\documentclass{article}
\usepackage[final]{neurips_2019}
\usepackage{longtable}
\usepackage{lscape}
\usepackage{booktabs}
\usepackage{amsfonts}
\usepackage{amsbsy}
\usepackage{tabu}
\usepackage{amsmath}
\numberwithin{figure}{section}

\usepackage[utf8]{inputenc} 
\usepackage[T1]{fontenc}    
\usepackage{hyperref}       
\usepackage{url}            

\usepackage{nicefrac}       
\usepackage{microtype}      
\usepackage{indentfirst}
\usepackage{algorithm}  
\usepackage{algorithmic} 
\usepackage{graphics}
\usepackage{graphicx}
\usepackage{float}
\usepackage{subfig}
\usepackage{verbatim}
\newcommand{\tabincell}[2]{
	\begin{tabular}{@{}#1@{}}#2\end{tabular}
}

\title{Statistical computation methods for microbiome compositional data network inference} 

\author{Liang Chen$^{1,\dag}$, Qiuyan He$^{1,\dag}$,  Hui Wan$^{1,\dag}$, Shun He$^{1}$, Minghua Deng$^{1,2,3,\ast}$\\
$^{1}$School of Mathematical Sciences, Peking University, Beijing, 100871, China\\
$^{2}$Center for Quantitative Biology, Peking University, Beijing, 100871, China\\
$^{3}$Center for Statistical Science, Peking university, Beijing, 100871, China\\
$^\ast$To whom correspondence should be addressed;\\
$\dag$ Equal contribution, in alphabetic order. \\
E-mail: dengmh@pku.edu.cn}

\begin{document} 

     \maketitle 

\textbf{keywords: }Microbiome compositional data, Network inference, Comprehensive review

\begin{abstract}
  Microbes can survive in some extreme environments and can be found almost everywhere in the world. Microbial communities have been found to be associated with higher live forms, including animals and plants. Microbes can affect processes from food production to human health, such as disease and homeostasis. Such microbes are not isolated, but rather interact with each other and establish connections with their living environments. Understanding these interactions is essential to an understanding of the organization and complex interplay of microbial communities, as well as the structure and dynamics of various ecosystems. A common and essential approach toward this objective involves the inference of microbiome interaction networks. Although network inference methods in other fields have been studied before, applying these methods to estimate microbiome associations based on compositional data will not yield valid results. On the one hand, features of microbiome data such as compositionality, sparsity and high-dimensionality challenge the data normalization and the design of computational methods. On the other hand, several issues like microbial community heterogeneity, external environmental interference and biological concerns also make it more difficult to deal with the network inference. \\
  In this paper, we provide a comprehensive review of emerging microbiome interaction network inference methods. According to various assumptions and research targets, estimated networks are divided into four main categories: correlation networks, conditional correlation networks, mixture networks and differential networks. Their scope of applications, advantages, as well as limitations, are presented in this review. Since real microbial interactions can be complex and dynamic, no unifying method has, to date, captured all the aspects of interest. In addition, we discuss the challenges now confronting current microbial associations study and future prospects. Finally, we highlight that the research in microbial network inference requires the joint promotion of statistical computation methods and experimental techniques. Codes of most methods introduced in this review will be collected in https://github.com/Qiuyanhe/Statistical-computation-methods-for-microbiome-compositional-data-network-inference. \\ 
\end{abstract}

\section{Introduction}
Microbes exist widely in the biosphere, from fertile soil to warm spring water and from hard rocks to thin atmosphere, and even in extreme environments, such as high-temperature deep-sea liquid vents, rocky crusts in the depths of the earth, and cold poles \citep{Huber2007, Pikuta2007}. Microbes play a vital role in the biochemical cycle of the earth. In the ecological cycle of the biosphere, microbes not only act as producers, but also as the main decomposers of organic matter, demonstrating their importance as a force driving the biochemical cycle \citep{Falkowski2008, Madsen2011}. Massine numbers of microbes live in the human body, and their total number is estimated to be 10 times that of human cells \citep{Micah2007}. They are distributed in various parts of the human body, such as skin, mouth, and intestines. Among them, the intestine has the largest number and variety of microbes \citep{Filyk2016}. Microbes often exist in the form of communities, and they affect their own environment through synergy, which, in turn, affects the composition and structure of the microbial community. The genomes of these microbes constantly interact with the human genome to regulate the metabolism of the host, which is closely related to human health. In recent years, many studies have emerged to explore the relationship between microbes and human health. For example, intestinal microbes are closely related to obesity \citep{Turnbaugh2009}, diabetes \citep{Qin2012}, high blood pressure \citep{Holmes2008}, cardiovascular diseases \citep{Wang2011}, AIDS \citep{Lozupone2013} and cancer \citep{Schwabe2013}. Meanwhile, people’s eating habits and lifestyles will also affect the structure of intestinal flora \citep{Wu2011}. The Human Microbiome Project (HMP) initiated by the American National Institutes of Health (NIH) uses metagenomics to study the relationship among human body surface, microbial flora and human health \citep{Turnbaugh2007, Huttenhower2012, Methe2012}.

The microbiome refers to the collection of all microbial species and their genetic information and functions that exist in a specific environment. It includes not only interactions among microbes in the environment, but also the interactions between microbes and other species and even environments. In recent years, with the rapid development of sequencing technology, researchers can extract genomic information of all microbes from microbial communities in the natural environment. More and more novel bacterial species have been discovered and microbial sequencing data have grown exponentially \citep{Proal2017}. Such sequencing makes it easier to study the correlations among microbial communities and between microbes and hosts. For example, 16S rRNA sequencing is a widely used high-throughput sequencing technology for microbiome studies. The 16S rRNA gene is the most common genetic marker to identify microbes \citep{Wooley2010}, and its advantages are mainly reflected in three aspects. First, the 16S rRNA gene exists in all microbes and has the same function in most organisms, except for eukaryotic cells. Second, the length of 16S rRNA gene can reach 1500bp, which can provide enough variation to identify the types of microbes \citep{Janda2007}. Third, 16S rRNA sequences have highly conserved regions among different species, which allow us to use universal primers to sequence multiple species at the same time and distinguish the type of microbes based on the given hypervariable regions \citep{Chakravorty2007}. Compared with other sequencing technologies, 16S rRNA sequencing technology also provides fast sequencing speed and low sequencing cost, which are also important factors in its widespread application in microbiome research. In the specific sequencing process, the variable regions of the 16S rRNA gene sequence are first amplified and sequenced, and then the sequences with similarity of 95\%, 97\% or 99\% are grouped into one Operational Taxonomic Unit (OTU); the OTU count is then used as proxy variable for potential microbial abundance \citep{Morgan2012, Schloss2010}. Furthermore, grouping 16S rRNA gene sequences to obtain OTUs can be achieved in three ways. The first is to cluster similar sequences using an unsupervised clustering algorithm \citep{Schloss2009}, and the second is to use a phylogenetic tree that integrates mutation and evolution information \citep{Hamady2010}. The third method is to use the supervised learning model obtained from the labeled training data to determine the OTU type corresponding to the sequences \citep{Wang2007}. This method is also applied to whole-genome shotgun sequencing. In order to control the experimental deviation caused, for example, by sequencing depth or biological sample size, the total count in the sample is typically used to normalize the count of each OTU \citep{Ozsolak2011, Ni2013, Fernandes2014}. These normalized data only reflect the relative abundance information of each OTU. In this way, microbiome sequencing data typically consist of compositional data \citep{SPIECEASI2015}.

A wide range of interactions takes place among microbes coexisting in nature. These interactions are vital to the earth's biochemical cycle and have a profound impact on human health and disease. Based on their influence on members of the microbial community, the various interactions among microbes can be divided into three types: beneficial, neutral and harmful \citep{Zengler2018, Lidicker1979, Faust2012}. Beneficial interactions include commensalism and mutualism. The former means that the interaction is beneficial to one microbe but has no effect on other microbes. The latter indicates that microbes coexist and benefit from each other. Neutral interaction represents the absence of any effect on both microbes. Harmful interactions are mainly divided into four situations: amenalism, competition, parasitism and predation. Amenalism means that the connection is harmful to one microbe but has no effect on other microbes. Competition implies that all microbes involved in the interaction are at a disadvantage owing to the existence of other microbes. Parasitism occurs when microbes live off of their host, benefiting themselves at the expense of their hosts. Predation means that one microbe feeds on another microbe. Together, these interactions constitute a complex ecological network which determines the structure, composition and function of the microbial community in the ecosystem.

In microbiology and systems biology research, probabilistic graph models are typically used to characterize the interactions between various microbes in a microbial community \citep{Whittaker2009, Jordan2003, Friedman2004, COAT2019}. We can regard each kind of microbe as a vertex. If an interaction occurs between two microbes, we can connect their corresponding vertices to form an edge. In this way, we can get a network model describing the interactive relationship between microbes, namely the microbial interaction network. Furthermore, we can determine the strength of the interaction between microbes by assigning weights to edges such that the greater the absolute value of the weight, the stronger the interaction. A positive weight indicates beneficial interaction, and a negative weight indicates harmful interaction. Zero weight means neutral interaction. In the probabilistic graphical model, the absolute abundance of each microbe is regarded as a random variable, making the absolute abundance of all microbes in the microbial community a random vector. Microbiome studies are concerned with how the interaction network corresponds to each component in this random vector. 

The goal of microbiome compositional data network analysis is to utilize the observed relative abundance data of various microbes in order to make statistical inferences about the interaction network corresponding to their absolute abundance. Such interactions are often inferred by estimating the covariance matrix. Approaches to these correlative estimations can be generally divided into Pearson correlation estimation and Lasso based regression. In detail, interactions can be divided into those that are direct and those that are indirect. Indirect interaction may be mediated between two microbes, or it may be transmitted through a third microbe. Direct interactions require no mediator or third factor and are more intrinsic. Since the two correlative estimation methods noted above do not make such distinctions, conditional correlation network inference methods that can estimate direct interactions have gained much attention. These approaches include graphical model inference, especially the Gaussian graphical model, and hierarchical model estimation based on different distribution assumptions. In addition, the interaction network between microbes is not constant for a long time, but changes as the external environment changes, such as the health status of the host, environmental conditions, or genetic conditions. Hence, research on differential networks and mixture networks has recently become an important topic in microbiome studies. Mixture networks account for the heterogeneity of microbiome data. Mixture networks methods generalize the single underlying biological network into multiple networks. They are usually constructed by mixing multiple distributions and solved by estimating parameters of hierarchical models. Differential networks focus on the dynamics of interaction networks and explore interactions from the aspect of comparison. Inference of hierarchical models can also help in estimating differential networks. Methods of estimating differential networks can also be classified as regression and Markov random fields.

Previously, several researchers have reviewed the inference methods of microbiome networks, but they mainly focused on correlation networks \citep{Weiss2016, Dohlman2019, Cardona2016}. Therefore, in this article, we review some of the statistical calculation methods developed in recent years based on all four main network situations. We discuss their model assumptions and scope of applications and provide some suggestions and prospects for future development of algorithms. Our paper is organized as follows. In the second section, we illustrate notations and some background knowledge. From the third to the sixth section, we introduce typical estimation methods of correlation networks, conditional correlation networks, mixture networks and differential networks accordingly. After that, we discuss existing challenges to microbiome interaction network inference and aspects which can be further studied in the future. All methods reviewed here are presented in Table \ref{table1:summary}.

\section{Notations and Background}
\subsection{Notations}
Before introducing methods, we illustrate some notations and background knowledge used in all subsections of this paper.

For any matrix $ A=(a_{ij}) $, $ \boldsymbol{a}_{i \cdot} $ and $ \boldsymbol{a}_{\cdot i} $ denote the $ i $-th row and column of $ A $, respectively. Besides, we use $ A_{(-j)} $ to denote the submatrix of $ A $ after removing its $ j $-th column. The determinant, trace and transposition of $A$ are denoted as $ \det(A) $, $ \text{tr}(A) $ and $ A^{T} $. Define $ \left\|A \right\|_{F}=( \sum_{i,j}a_{ij}^{2})^{\frac{1}{2}} $, $ \left\| A\right\|_{1} =\sum_{i,j}|a_{ij}| $ and $ \left\|A\right\|_{1,\text{off}}=\sum_{i\neq j}|a_{ij}| $. $ A_{1}\succeq A_{2} $ means that $ A_{1}-A_{2} $ is positive definite. The product $ <\cdot,\cdot> $ for two matrices represents $ <A_{1},A_{2}>=\text{tr}(A_{1}A_{2}^{T}) $. For any vector $ \boldsymbol{b} $, $ \left\|b \right\|_{k}=( \sum_{i}|b_{i}|^{k})^{\frac{1}{k}} $ is the $ \ell_{k} $-norm of $ \boldsymbol{b} $, $ k\in\mathbb{Z}^{+} $. The function $ s(\cdot) $ sums all elements of the vector $ \boldsymbol{b} $, that is, $ s( \boldsymbol{b} )=\sum_{i}b_{i} $. Moreover, $ \forall k\in\mathbb{Z}^{+} $, $ I_{d} $ and $ \boldsymbol{1}_{d} $ denote $ d $-dimensional identity matrix and all-ones vector, respectively.

In this paper, we denote the observed count matrix as $ Y=(y_{ij})_{n\times p} $ and the unobserved absolute abundance matrix as $ Z=(z_{ij})_{n\times p} $. Let 
\begin{equation}
x_{ij}=\frac{z_{ij}}{\sum_{j=1}^{p}z_{ij}},
\label{eq:Notations_relative}
\end{equation}
then the relative abundance matrix is $ X=(x_{ij})_{n\times p} $. The sample version of $ x_{ij} $ can be written as $ \frac{y_{ij}}{\sum_{j=1}^{p}y_{ij}} $. Besides, let $ W $ be the centered log-ratio (clr) transformed data \citep{Aitchison1981}. Specifically, for the $ i $-th sample, 
\begin{equation}
\begin{aligned}
\boldsymbol{w}_{i \cdot} \overset{\text{def}}{=} \text{clr}(\boldsymbol{x}_{i \cdot})
&=\left[\log\left(\frac{x_{i1}}{g(\boldsymbol{x}_{i \cdot})} \right),\dots,\log\left(\frac{x_{ip}}{g(\boldsymbol{x}_{i \cdot})} \right)  \right] \\
&=\left[\log\left(\frac{y_{i1}}{g(\boldsymbol{y}_{i \cdot})} \right),\dots,\log\left(\frac{y_{ip}}{g(\boldsymbol{y}_{i \cdot})} \right)  \right] \\
&=\left[\log\left(\frac{z_{i1}}{g(\boldsymbol{z}_{i \cdot})} \right),\dots,\log\left(\frac{z_{ip}}{g(\boldsymbol{z}_{i \cdot})} \right)  \right] ,
\end{aligned}
\label{eq:Notations_clr}
\end{equation}
where $ g(\boldsymbol{x}_{i \cdot})=( \prod_{j=1}^{p}x_{ij})^{1/p} $, $ g(\boldsymbol{y}_{i \cdot})=( \prod_{j=1}^{p}y_{ij})^{1/p} $ and $ g(\boldsymbol{z}_{i \cdot})=( \prod_{j=1}^{p}z_{ij})^{1/p} $. The clr transformation is isometric and symmetric with regard to components. Denote the covariance of log-transformed relative abundances in a sample as $ \Omega=\text{cov}(\log \boldsymbol{x}) $, where $ \boldsymbol{x}=(x_{1},\dots,x_{p}) $. Then denote the covariance of log-transformed absolute abundances in a sample as $ \Sigma=\text{cov}(\log \boldsymbol{z}) $ and its inverse matrix (precision matrix) as $\Theta$, where $ \boldsymbol{z}=(z_{1},\dots,z_{p}) $. Let $ \Gamma=\text{cov}(\boldsymbol{w})=\text{cov}(\text{clr}(\boldsymbol{x}))=\text{cov}(\text{clr}(\boldsymbol{z}))$, where $ \boldsymbol{w}=(w_{1},\dots,w_{p}) $. Based on the simple calculation in \citep{Aitchison1986}, we have 
\begin{equation}
\label{eq:Notations_clrBridge}
\Gamma=F\Sigma F,
\end{equation}
where $ F=I_{p}-\frac{1}{p}\boldsymbol{1}_{p}\boldsymbol{1}_{p}^{T} $. Besides, using Equation~\ref{eq:Notations_relative} and $ F\boldsymbol{1}_{p}=\boldsymbol{0}_{p} $, we can get
\begin{gather}
\label{eq:Notations_log_relative}
\log\boldsymbol{x}=\log\boldsymbol{z}-\boldsymbol{1}_{p}\log s(\boldsymbol{z}), \\
F\log\boldsymbol{x}=F\log\boldsymbol{z}.
\label{eq:Notations_Flog}
\end{gather}
Then by calculating variance on two sides of Equation~\ref{eq:Notations_Flog}, we get a new equation:
\begin{equation}
F\Omega F=F\Sigma F.
\label{eq:Notations_varBridge}
\end{equation}

If assuming  that  the  count  data $ \boldsymbol{y}_{i\cdot} \sim \text{Multinomial}(\boldsymbol{\alpha}_{i\cdot}),$ where $ \sum_{j=1}^{p}\boldsymbol{\alpha}_{ij}=1 $,  additive log-ratio (alr) transformation to $ \boldsymbol{\alpha}_{ij},i=1,\dots,n$ is defined as:
\begin{equation}
	\boldsymbol{h}_{i\cdot}=\left(\log\left(\frac{\alpha_{i1}}{\alpha_{ip}}\right),\dots, \log\left(\frac{\alpha_{i,p-1}}{\alpha_{ip}}\right)\right)  ,i=1,\dots,n.
	\label{eq:Notation_alr}
\end{equation}

In this review, $ M=(m_{ij})_{n\times q} $ denotes environmental factor matrix, and $ m_{ij} $ is the value of the effect that the $ j $-th factor has on the $ i $-th sample.

\subsection{Features of compositional data}
Sequencing depth is defined as the total sequencing count summed over all taxa of a sample and can vary across samples. Such variation of sequencing depth may be mainly caused by experimental techniques. Considering this fact, the microbial count data only carries information of relative abundances instead of absolute abundances. That is the compositionality of microbiome data. When the count value of a taxon increases in a sample, the relative abundances of all other taxa in this sample decrease, even if their absolute abundances do not change. Owing to the sum-constant constraint, traditional correlation statistics may give spurious results. Take Pearson correlation as an example:
\begin{equation}
	\sum_{j=1}^{p}x_{ij}=1 \Rightarrow 
	\sum_{j\neq k}\text{cov}(x_{ij},x_{ik}) = -\text{var}(x_{ik}).
\end{equation}
Thus the correlation is characterized by a negative trend, even though the true abundances may be independent. Such problem can be managed with by ratio transformation such as clr and alr transformations introduced above.

Moreover, the number of features in the microbiome data is usually far larger than the sample-size. In another words, microbiome data are high-dimensional. This high-dimensionality can lead to heavy computational burden. Algorithms need to be designed carefully for efficient computation. Since the sequencing technique keeps improving, the scalability onto high-dimensional data having larger-scale is necessary.

Another feature of microbiome data is excessive zeros. This poses challenges to data transformation. Many methods add a pesudo-count to avoid invalid computation. The choice of pesudo-count is subjective and may affect the performance of the method. In fact, zeros in the count data can be generally divided into two types. One type results from taxa are present, but in proportions too low to be detected, and such zeros are often called technical zeros. Another type caused by the true absence of certain taxa, leading to essential zeros, is also called structural zeros \citep{NetCoMi2020}. Some methods choose to model these zeros, but the performance depends on the rationality of distribution assumptions.

The composition of microbial communities can drastically vary from hosts to environments \citep{Jiang2019}. Thus when collecting data of many samples, a taxon that is relatively abundant in one sample may not be observed in other samples. This is the sparsity mentioned above. Another phenomenon is the tremendous variation among the dominant taxa collected in just one sample. This is termed as taxonomic heterogeneity by \citep{Jiang2019}. In addition to taxonomic heterogeneity, multiple underlying microbial networks may determine the sample-taxa matrix in reality, rather than one single network. For example, in the human gut, the interaction between two taxa may depend on the enterotypic context of the individual's gut microbiome that can vary from individual to individual \citep{Jiang2019}. This is another type of heterogeneity which is functional.

\section{Correlation network}
In order to identify the interactions among members of natural microbial communities, correlation analysis is proposed to measure the dependencies among microbes. However, the observed data only provide relative fractions of species. For compositional data, standard correlation analytic methods could result in spurious correlations. Therefore, various specialized methods based on compositional data have been developed to analyze the microbiome correlation network. 

SparCC \citep{SparCC2012} is just about the earliest method that takes compositionality into consideration. It employs the log-ratio transformation to estimate Pearson correlations between components. SparCC also has an iterative version that identifies component pairs having high correlations at each step to guarantee sparsity until the recognition of strongly correlated pairs is stable. Another method is regression-based and utilizes $ \ell_{1} $ penalty to estimate sparse covariance matrix, such as CCLasso \citep{CCLasso2015} and REBACCA \citep{REBACCA2015}. CCLasso follows SparCC to perform on log-ratio transferred compositional data. CCLasso guarantees a positive correlation matrix, but both CCLasso and SparCC rely heavily on the quality of component fraction estimations. The performance of CCLasso may not be satisfying if the number of taxa is much larger than that of samples. REBACCA builds a linear system which is equivalent to the log-ratio transformations. REBACCA may also fail when sample size is small and it can be strongly affected by noise. 

Both CCLasso and REBACCA lack theoretical guarantees, as well as the required explicit statistical assumptions \citep{COAT2019}. So, COAT \citep{COAT2019} was developed to estimate the sparse covariance matrix in a principled way and does provide theoretical guarantees on support recovery. It shows that the basis covariance is approximately identifiable if it belongs to the large sparse covariance family \citep{COAT2019}. COAT is equivalent to thresholding the sample-centered covariance matrix after log-ratio transformation, and it can be applied to large-scale data. However, all methods mentioned above can only estimate linear correlation relationships.

\subsection{SparCC}
SparCC (Sparse Correlations for
Compositional data) \citep{SparCC2012} introduces correlation between log transformed absolute abundances based on compositional data. It estimates the linear Pearson correlations between the log-transformed components \textit{via} approximation based on assumptions that the number of different OTUs is large and the actual correlation network is sparse. It proposes a basic procedure to infer such linear Pearson correlations and uses an iterative procedure to improve the result. Specific steps can be summarized as follows:

The basic inference procedure defines the covariance matrix $T=(t_{ij})_{p\times p}$ containing information about the dependencies between components as:
\begin{equation}
	\begin{aligned}
		t_{ij}&\overset{\text{def}}{=}\text{var}\left(\log\frac{x_{i}}{x_{j}} \right)=\text{var}\left(\log\frac{z_{i}}{z_{j}} \right)  \\
		&=\text{var}\left(\log z_{i}\right)+\text{var}\left(\log z_{j}\right)-\text{cov}\left(\log z_{i},\log z_{j}\right) \\
		&=\sigma_{ii}+\sigma_{jj}-2\rho_{ij}\sqrt{\sigma_{ii}\sigma_{jj}},
	\end{aligned}
	\label{eq:SparCC_Tdef}
\end{equation}
in which $ \rho_{ij} $ is the correlation of the log transformed absolute abundance variable $ i $ and $ j $, and $ \sigma_{ij}=\text{cov}(\log z_{i},\log z_{j}) $. SparCC also adopts a bayesian framework to get the full posterior joint distribution of fractions, as well as their point estimated, in order to obtain the sample version of $ T $. Since SparCC assumes that the average correlations are small, all basis variances $ \sigma_{jj},j=1,\dots,p $ can be estimated by solving the approximate set of equations:
\begin{equation}
	t_{i}=(p-1)\sigma_{ii}+\sum_{j\neq i}\sigma_{jj},\quad i=1,\dots,p.
\end{equation}
Then the estimated correlations can be achieved \textit{via} Equation~\ref{eq:SparCC_Tdef}:
\begin{equation}
	\rho_{ij}=\frac{\sigma_{ii}+\sigma_{jj}-t_{ij}}{2\sqrt{\sigma_{ii}\sigma_{jj}}}.
\end{equation}
After defining the basic inference procedure, iterative SparCC identifies strongly correlated component pairs if the correlation is larger than the threshold chosen, and removes components of such pair until no new strongly correlated pair is left, or only three components remain. The above step is used to guarantee the sparsity assumption. Once any component is removed, the component fractions are re-estimated and so are the component variations.

In fact, SparCC estimates the correlations of log transformed absolute abundances without the involvement of relative information. Although absolute abundances can not be observed, SparCC can be free of compositional bias using this concept, implying that SparCC relies heavily on the quality of component counts data. Also, SparCC adds pseudo-counts to eliminate zero fractions, which may affect the correlation estimations. SparCC is biased towards positive correlations and does not perform well given highly diverse samples. While diversity of species influences the severity of compositional effects on correlation estimation, SparCC assumes sparsity rather than relying on high diversity. However, these approximate assumptions induce errors during the solution process that may influence accuracy. Moreover, the inferred covariance matrix can not be guaranteed positive definite, and correlation coefficients may not locate in $ [-1,1] $. SparCC may not be computationally efficient with high species dimension, and as pointed out in \citep{SparCC2012}, it only estimates linear relationships between log-transformed abundances.
    
\subsection{CCLasso}
Similar to SparCC, CCLasso (Compositional data Correlation inference through Lasso) \citep{CCLasso2015} considers the compositional property of the observed data in correlation analysis explicitly, and it has an extra advantage in that the estimated correlation matrix of the latent variables for compositional data is positive definite. CCLasso estimates correlations of log-transformed absolute abundance \textit{via} weighted least squares with $ \ell_{1} $ penalty based on log-transformed relative abundance. In order to solve the optimization problem, it proposes an alternating direction algorithm from augmented Lagrangian method.
    
Choosing a matrix $F=I_{p}-\frac{1}{p}\boldsymbol{1}_{p}\boldsymbol{1}_{p}^{T} $ for clr transformed data \citep{Aitchison1986} satisfying $ \text{rank}(F)=p-1 $ and $ F\boldsymbol{1}_{p}=\boldsymbol{0}_{p} $, as proved in Equation \ref{eq:Notations_varBridge}, the variance of $F\log \boldsymbol{x}$ equals that of $F\log \boldsymbol{z}$.
The rank requirement of $ F $ makes $ F\log \boldsymbol{x} $ correspond to $ \boldsymbol{x} $ one-to-one. The estimation of fractions (compositions) can lead to the sample version of $ \Omega$, denoted as $ S $, and CCLasso tries to measure the distance (or discrepancy) between $ F\Sigma F^{T} $ and $ FSF^{T} $. Inspired by \citep{Dtrace2014} and as $ F$ is idempotent and symmetric, CCLasso then proposes the following loss function:
\begin{equation}
    \begin{aligned}
    \mathcal{L}\left(\Sigma\right)&=\frac{1}{2}\text{tr}\left( \left(F\left(\Sigma-S \right)F^{T}\right)V\left(F\left(\Sigma-S \right)F^{T}\right)\right) \\
    &=\frac{1}{2}\left\|F\left(\Sigma-S \right)F\right\|_{V}^{2} . 
    \end{aligned}
\end{equation}
Besides, CCLasso incorporates the sparse assumption by adding $ \ell_{1} $ penalty on the off-diagonal elements of $ \Sigma $. Then the optimization problem is proposed as:
\begin{equation}
    \widehat{\Sigma}=\mathop{\arg\min}_{\Sigma\succ 0}\frac{1}{2}\left\|F\left(\Sigma-S \right)F\right\|_{V}^{2}+\lambda \left\|\Sigma \right\|_{1,off}. 
    \label{eq:CCLasso_RAWopt}
\end{equation}
Correlation matrix estimated from such optimization problem is positive definite and each element lies in $[-1,1] $, which can not be guaranteed by SparCC.
    
To solve the constrained optimization problem in Equation~\ref{eq:CCLasso_RAWopt},  \citep{CCLasso2015} proposes an alternating direction method to minimize the augmented Lagrangian function.

However, similar to the limitations of SparCC, CCLasso needs reliable component fraction estimation, only explains linear relationships and doesn't offer theoretical performance guarantees. In terms of managing excessive zeros, CCLasso also adops a simple pseudo count to avoid zero components, which may affect its performance. 
    
\subsection{REBACCA}
REBACCA (Regularized Estimation of the Basis Covariance based on Compositional data) \citep{REBACCA2015} identifies important co-occurrence patterns by finding sparse solutions to a system with a deficient rank. Different from SparCC, REBACCA constructs a linear system that is equivalent to log-ratio transformations and solves the system using the $\ell_1$-norm shrinkage method under a sparsity assumption.

Let $ t(\boldsymbol{z})=\sum_{i\neq j}\text{var}\left(\log\frac{z_{i}}{z_{j}}\right)$
and denote $ \boldsymbol{z}_{-\{k,l\}} $ as the subvector of $ \boldsymbol{z} $ after deleting the $ k $-th and $l$-th elements. 
Denote $u_{kl}= \frac{t(\boldsymbol{z})}{2(p-1)}-\frac{t(\boldsymbol{z}_{(-l)})}{2(p-2)}-\frac{t(\boldsymbol{z}_{(-k)})}{2(p-2)}+\frac{t(\boldsymbol{z}_{-\{k,l\}})}{2(p-3)}$.
Range the upper diagonal part of $ \Sigma $ as $ \boldsymbol{v}=(\sigma_{1p},\sigma_{2p},\dots,\sigma_{p-1,p},\sigma_{1,p-1},\\\dots,\sigma_{13},\sigma_{23},\sigma_{12})^{T} $, and denote $ \boldsymbol{a}_{kl}^{T} $ as the $ \frac{p(p-1)}{2} $ coefficient vector of $ \boldsymbol{v} $, where $u_{kl}=\boldsymbol{a}_{kl}^{T}\boldsymbol{v}$. Considering all possible pairs, let $ \boldsymbol{u}=(u_{p1},u_{p2},\dots,u_{p,p-1},u_{p-1,1},\\\dots,u_{21})^{T} $ and $ A^{T}=(\boldsymbol{a}_{p1},\boldsymbol{a}_{p2},\dots,\boldsymbol{a}_{p,p-1},\boldsymbol{a}_{p-1,1},\dots,\boldsymbol{a}_{21})^{T} $. Then, the linear system after some calculations can be obtained:
\begin{equation}
    \boldsymbol{u}=A\boldsymbol{v}.
    \label{eq:REBACCA_linearEq}
\end{equation} 
REBACCA incorporates $ \ell_{1} $ penalty onto off-diagonal elements of $ \Sigma $ since the number of variables is larger than $ \text{rank}(A) $. Equation~\ref{eq:REBACCA_linearEq} has a unique solution under some specific conditions. REBACCA solves it by least squares, which can be written as a Lasso problem:
\begin{equation}
    \min_{\boldsymbol{v}}\frac{1}{2}\left\|A\boldsymbol{v}-\boldsymbol{u}\right\|_{2}^{2} + \lambda\left\|\boldsymbol{v} \right\|_{1}.
    \label{eq:REBACCA_lasso} 
\end{equation}
    
Then REBACCA proposes a stability resampling method to control the expected number of variables selected with low probability at a given significance level, and, lastly, $ \sigma_{ii},i=1,\dots,p $ can be calculated \textit{via} equations: $ \text{var}(\log\frac{z_{i}}{z_{j}})=(\sigma_{ii}+\sigma_{jj}-2\sigma_{ij}) ,i,j=1,\dots,p$.

Similar to SparCC, REBACCA estimates pair-wise relationships of log transformed absolute abundances, so it's free of compositional bias. However, the iterative process may not scale to $ p $, resulting in high computational requirement. Besides, no theoretical performance guarantees are provided for REBACCA to work effectively.

\subsection{COAT}
In order to solve the problems existing in the above methods and provide statistical insights into the issue of identifiability and the impacts of dimensionality, \citep{COAT2019} addresses the problem of covariance estimation for high-dimensional compositional data. Under the assumption that the basis covariance matrix is sparse, a composition-adjusted thresholding (COAT) estimator is defined.

COAT is based on a decomposition relating the centered log-ratio covariance matrix $\Gamma$ to the basis covariance matrix $\Sigma$, which is approximately identifiable as the dimensionality goes to infinity. Hence, \citep{COAT2019} first calculates the estimate of $ \Gamma$, which is $ \widehat{\Gamma}=(\hat{\gamma}_{ij})_{p\times p} $  \textit{via} clr transformed data $ W=(w_{kj})_{n\times p}$, satisfying:
\begin{equation}
\hat{\gamma}_{ij}=\frac{1}{n}\sum_{k=1}^{n}\gamma_{kij}=\frac{1}{n}\sum_{k=1}^{n}\left(w_{ki}-\bar{w}_{i}\right)\left(w_{kj}-\bar{w}_{j}\right),
\end{equation}
where $\bar{w}_{j}=\frac{1}{n}\sum_{k=1}^{n}w_{kj} $.

Second, \citep{COAT2019} applies adaptive thresholding $ S_{\delta}(\cdot) $, satisfying certain conditions to $ \widehat{\Gamma} $ to define the composition-adjusted thresholding (COAT) estimator $ \widehat{\Sigma}=(\sigma_{ij})_{p\times p} $:
\begin{equation}
	\hat{\sigma}_{ij}=S_{\delta_{ij}}\left(\hat{\gamma}_{ij}\right),
\end{equation}  
where the data-driven $ \delta_{ij}=\lambda\sqrt{\hat{\xi}_{ij}} $ and $\hat{\xi}_{ij}=\frac{1}{n}\sum_{k=1}^{n}\left( \gamma_{kij}-\hat{\gamma}_{ij}\right)^{2}$.

Furthermore, Cao et al. \citep{COAT2019} also establish rates of convergence and provide support recovery guarantees for this estimator.

COAT is equivalent to thresholding the sample centered log-ratio covariance matrix. So, it is free of optimization and efficient even when $ p $ is large. This work also fills the gap of large covariance estimation on constrained data by adopting covariance thresholding methods to compositional data. The work of COAT includes explicit statistical assumptions needed for it to be efficient, and guarantees its performance theoretically. Of course, the estimated correlation relationship is linear, and COAT can-not handle excessive zeros directly.

\section{Conditional correlation network}
Taking the compositionality of microbiome data into consideration, a microbial interaction network describes the pairwise correlations between two taxa. To restate, such correlations can be divided into direct interactions and indirect interactions. Indirect interactions involve two taxa correlated by some confounding factor, such as a certain environment condition, or a third species. Direct interaction means the nonexistence of confounding factors, and one taxon influences another directly. Since direct interactions are usually more fundamental and intrinsic \citep{Friedman2004}, methods modeling the conditional dependencies among taxa given all other taxa have gained much attention.

To infer such conditional dependencies, many methods turn to estimating a probabilistic graphical model in which each node represents a taxon, and each edge is undirected, indicating the conditional dependency of two taxa \citep{Codaloss2020}. If data follow a distribution in the exponential family, a straightforward distributional interpretation of such models can be made \citep{WainwrightJordan2008}. Taking the multivariate Gaussian distribution as an example, the element of the inverse covariance matrix is zero if and only if its two corresponding variables are conditionally independent of all other variables. Therefore, to infer the conditional correlation network where an edge between two nodes represents conditional dependency given all other nodes, many researchers turn to estimating the non-zero entries of the inverse covariance matrix. The inverse covariance matrix is also termed as precision matrix. For real microbiome data, which are high-dimensional and sparse, the calculated covariance matrix is usually degenerate. Thus, estimating the precision matrix is challenging. 

SPIEC-EASI \citep{SPIECEASI2015} estimates the covariance of the log-ratio transformed absolute abundance approximately by calculating that of centered log-ratio transformed data. Its overall inference procedure includes estimating graphical models, selecting neighborhoods and sparse inverse covariance. However, it's difficult for SPIEC-EASI to recover networks having large hubs and the recovery performance is affected by the condition number of the inverse covariance. gCoda \citep{gCoda2017} assumes that the absolute abundances follow a multivariate log-normal distribution and that the underlying network is sparse. Therefore it can model the generation of compositional abundance by a logistic normal distribution, and then estimate the network by maximum likelihood, using $ \ell_{1} $-penalty to ensure sparsity. Compared with SPIEC-EASI, gCoda is more computionally efficient and stable. MPLasso \citep{MPLasso2017} uses biological knowledge mined the literature as prior information in the graphical Lasso. The prior information is utilized to restrict the search space and make interaction inference more reliable. 

CD-trace loss is a novel empirical loss function proposed in \citep{CDTrace2019} based on D-trace \citep{Dtrace2014} loss, which is used to estimate high dimensional sparse precision matrices. CD-trace loss is convex, making the numerical solution easier to obtain when compared to gCoda. The bridge in Equation~\ref{eq:Notations_varBridge} between unobserved variables and observed compositional data is utilized by CD-trace to estimate the transformed covariance matrix. Similar to gCoda, the consistency of estimators obtained by CD-trace can not be guaranteed. CDTr \citep{CDTr2019} is another loss function based on D-trace \citep{Dtrace2014} loss. It incorporates clr transformation to infer the precision matrix and considers the compositionality of data. CDTr has a more concise form than CD-trace loss, but it still can not guarantee the consistency of results. Moreover, the exchangeable condition used in CD-trace and CDTr needs theoretical verification and its biological reasonableness needs further checking. 

MInt \citep{MInt2016} proposes a Poisson-multivariate normal hierarchical model to capture direct interactions. Confounding factors are controlled at the Poisson layer and the multivariate normal layer to estimate interactions among taxa. The Poisson-multivariate normal model has a flexible mean-variance relationship and can manage overdispersion conveniently. However, this model directly handles count data in a manner that does not consider the compositionality of microbiome data explicitly. mLDM \citep{mLDM2017} proposes a hierarchical Bayesian model to infer direct ineractions between microbes, as well as associations between microbes and environmental factors. mLDM considers the compositional bias and variance of the data. However, mLDM contains so many interim parameters that the computation burden is heavy when sample-size is large. Thus the scalability and efficiency of mLDM are limited. 

As is analyzed before, microbiome data have the features of compositionality, sparsity and high-dimensionality. Applying log-ratio transformation onto compositional data can free it from the unit-sum constraint, as well as preserve the counts ordering in each sample. Meanwhile, however, it cannot distinguish among different types of zeros in composition data. Sometimes such confusion from different types of zeros may cause bias in network estimation. 

HARMONIES \citep{HARMONIES2020} can meet the challenge of this problem by normalizing data using the zero-inflated negative binomial (ZINB) model with the Dirichlet process as a prior (DPP). The ZINB model can capture zero counts and over-dispersion. The Dirichlet prior can take sample heterogeneity into consideration. After normalization, HARMONIES estimates the partial correlations to infer direct associations among taxa. To avoid the bias of estimators caused by proportion-based methods, BC-GLASSO \citep{BCGLASSO2020} directly models compositional count data by a multinomial distribution. Then it assumes that multinomial probabilities follow a logistic normal distribution. The procedure of BC-GLASSO is similar to that of SPIEC-EASI, but it defines the true covariance matrix based on multinomial probabilities after the additive log-ratio (alr) transformation rather than abundances after clr transformation. Then BC-GLASSO conducts a bias-correction procedure for the estimator of true covariance matrix, during which the unevenness of sequencing depths can be handled. Based on this bias-corrected estimator, the graphical lasso is applied to infer a sparse inverse covariance matrix. The compositional graphical lasso proposed in \citep{CompGlasso2020} also adopts the logistic normal multinomial distribution to model the count data, which is two-level hierarchical. However, the objective function used in it is derived from log-likelihood which is different from that of BC-GLASSO. The theoretical convergence of the algorithm is established in \citep{CompGlasso2020}, while the theoretical property of the estimator remains to be elucidated. Alr transformation involves choosing a reference taxon which may affect network estimation. Therefore, \citep{InvGlasso2020} further analyses are needed to determine if using a different reference taxon could change the nature of the estimated network in compositional graphical lasso and further affect its robustness. The reference-invariance property established in \citep{InvGlasso2020} shows that the specific submatrix of the precision matrix is invariant to the choice of reference taxon. According to this property, the reference-invariant version of compositional graphical lasso, which only penalizes the invariant submatrix, is proposed by \citep{InvGlasso2020}.

It can be easily seen that many methods use the Gaussian graphical model to infer conditional microbial interactions and that such model is combined with log-ratio transformation in many cases. On the one hand, the centered log-ratio transformation can-not directly manage abundance data with excessive zeros. On the other hand, the transformed data may not follow Gaussian distribution. Thus, fitting Gaussian graphical models onto it may not be suitable. Besides, it is difficult to interpret such models, and the graphical models discussed here do not assign statistical significance or uncertainties onto the estimated conditional correlations \citep{MarRanField2019}.

\subsection{SPIEC-EASI}
SPIEC-EASI (SParse InversE Covariance Estimation for Ecological ASsociation Inference) \citep{SPIECEASI2015} infers microbial ecological networks from compositional data which could be potentially high-dimensional. It focuses on direct interactions and estimates conditional independence by inferring an underlying graphical model.

\citep{SPIECEASI2015} transforms observed OTU count data $ Y $ into relative abundance matraix $ X $ at first. After adopting clr transformation \citep{Aitchison1981}, the covariance matrix $\Gamma$ can be tied to the covariance matrix $\Sigma$ of log-transformed absolute data through Equation~\ref{eq:Notations_clrBridge}. When $ p\gg 0 $, we find that $ F $ approximates the identity matrix. Hence for high-dimensional data, the empirical covariance $ \widehat{\Gamma} $ can be a good approximation of $ \widehat{\Sigma} $ \citep{SPIECEASI2015}.

After obtaining $ \widehat{\Gamma} $, \citep{SPIECEASI2015} tries to build a probabilistic graphical model \citep{Koller2009} to capture the conditional dependence structure of the microbiome network. The network is presented as an undirected, weighted graph $ \mathcal{G}=(V,E) $ in which a node represents a taxon, and the edge between two taxa means they are associated. Here, $ V $ is the node set and $ E $ is the edge set. To infer the graph structure, SPIEC-EASI adopts two ways.

The first way is neighborhood selection. Following the MB method proposed in \citep{MB2006}, SPIEC-EASI solves a $ \ell_{1} $-penalized linear regression problem for each node. Then the local conditional independence structure of each taxon can be mined. For any node $j$, the regression problem is:
\begin{equation}
\widehat{\boldsymbol{\beta}}^{j,\lambda}=\mathop{\arg\min}_{\boldsymbol{\beta}\in\mathbb{R}^{p-1}} \frac{1}{n}\left\|\boldsymbol{w}_{\cdot j}-W_{(-j)}\boldsymbol{\beta} \right\|_{2}^{2}+\lambda\left\|\boldsymbol{\beta} \right\|_{1},\  \lambda\geq 0 .
\end{equation} 
This is, indeed, a LASSO problem \citep{Tibshirani1996} which is convex. After fitting the regression, the neighborhood of the $ j $-th node is comprised of the nodes, the corresponding index of which is $\widehat{\boldsymbol{\beta}}^{j,\lambda}\neq 0$. Then the edge set of graph $ \mathcal{G} $ can be defined through intersecting or uniting the local neighborhood.

The second way is covariance selection, which is inspired by graphical LASSO \citep{El2008} under standard Gaussian distribution setting. It is solved through Equation~\ref{eq:SPIEC-EASI} which maximizes penalized likeihood given clr-transformed data $ W $. When the data $ \log \boldsymbol{z} $ are drawn from multivariate normal $ \mathcal{N}(\log \boldsymbol{z}|\boldsymbol{\mu},\Sigma) $, note that the precision matrix $ \Theta $ defines the adjacency matrix of graph $ \mathcal{G} $.
\begin{equation}
	\label{eq:SPIEC-EASI}
\widehat{\Theta}=\mathop{\arg\min}_{\Theta\succ 0,\Theta=\Theta^{T}}-\log\det\left(\Theta\right) +\text{tr}\left(\Theta\widehat{\Gamma} \right)+\lambda \left\|\Theta \right\|_{1}.
\end{equation}

It's worth noting that SPIEC-EASI scales as the dimension of data grows and can recover the graph even when $ p\gg n $. Although not specifically introduced here, \citep{SPIECEASI2015} also proposes a pipeline of generating synthetic data.

SPIEC-EASI combines clr transformation with a graphical model inference framework that assumes sparsity of the underlying ecological association network and that its inference engine has the advantage of incorporating prior knowledge about the underlying data or network structure from independent scientific studies in a principled manner. The key assumption holds that when $ p $ is large enough, $\hat{\Gamma}$ approximates $\hat{\Sigma}$, which relies heavily on the conditional number of the inverse covariance matrix. SPIEC-EASI also relies on the assumption that the proportion of a taxon in a sample approximates the true value, ignoring the uncertainty of such estimates. Moreover, it is difficult for clr transformation to handle excessive zeros, and the transformed data are not even close to obeying Gaussian or sub-Gaussian graphical models. Unique characteristics observed in count data may be taken into account for improvement. What's more, the consistency of estimated covariance matrix can not be guaranteed. 

\subsection{gCoda}
\citep{gCoda2017} uses a logistic normal distribution to model the observed microbiome data and proposes a method called gCoda based on maximum likelihood with $\ell_1$-penalty to estimate conditional dependence structures which indicate direct interactions in microbial communities.

At first, \citep{gCoda2017} assumes that $ \log \boldsymbol{z}_{i \cdot}\sim \mathcal{N}(\boldsymbol{\mu},\Sigma), i=1,\dots,n $ are independent and that, according to \citep{Aitchison1980}, the random compositional vectors $ \log \boldsymbol{x}_{i \cdot}, i=1,\dots,n $ follow a logistic normal distribution. After some calculations, the marginal distribution of $ \boldsymbol{x}_{i \cdot} $ can be obtained as:
\begin{equation}
p(\boldsymbol{x}_{i \cdot})=\frac{1}{(2\pi)^{\frac{p-1}{2}}}\left[\frac{\det\left(\Theta\right)}{\boldsymbol{1}_{p}^{T}\Theta\boldsymbol{1}_{p}} \right]^{\frac{1}{2}}\prod_{j=1}^{p}\frac{1}{x_{ij}}\exp\left(-\frac{Q}{2} \right),
\end{equation}
where $ Q=(F\log\boldsymbol{x}_{i \cdot}-F\boldsymbol{\mu})^{T}\left(\Theta-\frac{\Theta\boldsymbol{1}_{p}\boldsymbol{1}_{p}^{T}\Theta}{\boldsymbol{1}_{p}^{T}\Theta\boldsymbol{1}_{p}}\right)(F\log\boldsymbol{x}_{i \cdot}-F\boldsymbol{\mu}) $, and $ F=I_{p}-\frac{1}{p}\boldsymbol{1}_{p}\boldsymbol{1}_{p}^{T} $, as indicated in Notations.

Using the negative log likeihood as loss function, replacing $ \boldsymbol{\mu} $ with the empirical estimator $ \widehat{\boldsymbol{\mu}}=\frac{1}{n}\sum_{i=1}^{n}\log\boldsymbol{x}_{i \cdot} $, and assuming that the direct interaction network is sparse by adding $ \ell_{1} $-penalty, the optimization objective can be written as:
\begin{equation}
\widehat{\Theta}=\mathop{\arg\min}_{\Theta\succ 0}-\log\left[\frac{\det\left(\Theta\right)}{\boldsymbol{1}_{p}^{T}\Theta\boldsymbol{1}_{p}} \right]+\text{tr}\left[S\left(\Theta-\frac{\Theta\boldsymbol{1}_{p}\boldsymbol{1}_{p}^{T}\Theta}{\boldsymbol{1}_{p}^{T}\Theta\boldsymbol{1}_{p}}\right)\right]+\lambda \left\|\Theta \right\|_{1},
\end{equation}
where $ S=\frac{1}{n}\sum_{i=1}^{n}(F\log\boldsymbol{x}_{i \cdot}-F\widehat{\boldsymbol{\mu}})(F\log\boldsymbol{x}_{i \cdot}-F\widehat{\boldsymbol{\mu}})^{T} $.
However, as this optimization function is neither convex nor smooth, \citep{gCoda2017} proposes an efficient algorithm based on the MM principle \citep{Hunter2004, Lange2016} to solve the constrained optimization problem. 

gCoda is just about first method to derive the likelihood of compositional data for Gaussian graphical models, although the likelihood function is non-convex. Besides, the consistency of the estimator can not be guaranteed.

\subsection{MPLasso}
Microbial Prior Lasso (MPLasso) \citep{MPLasso2017} integrates the graph learning algorithm with microbial co-occurrence patterns and associations directly from a large amount of scientific literature by using automated text mining.

MPLasso tries to use an undirected, weighted graph $ \mathcal{G} $ to represent the pairwise associations between taxa. First, MPLasso  \citep{MPLasso2017} assumes that observed data follows a multivariate normal distribution $\mathcal{N}(\boldsymbol{\mu},\Sigma)$. However, since compositional data are usually high-dimensional, the graph inference problem is intractable. To cope with this problem, MPLasso \citep{MPLasso2017} assumes that the true graph is sparse. However, instead of adding $ \ell_{1} $-penalty on the precision matrix $ \Theta=\Sigma^{-1} $ directly, MPLasso incorporates the prior information onto the penalty matrix $ P=(p_{ij})_{p\times p} $. Hence the optimization function is written as:
\begin{equation}
\max_{\Theta}\log\det\left(\Theta\right)-\text{tr}\left(\Theta\widehat{\Gamma} \right)-\lambda\left\|P\odot\Theta \right\|_{1},   
\end{equation} 
in which $ \widehat{\Gamma} $ is the empirical covariance of the clr transformed data, and the operation $ \odot $ is the Hadamard product. When $ p_{ij} $ is large, the penalty is heavy, and the association between taxon $ i $ and $ j $ is weak. 

To obtain the prior information $ P $, \citep{MPLasso2017} utilizes the PubMed database which contains large numbers of research studies with abstracts. The mining method is performed according to the taxonomic level. For 16S rRNA data where the taxonomic level is often genus level, \citep{MPLasso2017} uses Fisher's exact test \citep{Freilich2010} to examine the microbial co-occurrence in the scientific literature. For shotgun data where the taxonomic level can be the species level, \citep{MPLasso2017} utilizes both the microbial co-occurrence in literature and the machine learning method proposed in \citep{Lim2016} to classify abstracts. 

MPLasso proposes to use experimentally verified biological knowledge as prior knowledge \textit{via} text mining to enhance accuracy in inferring associations. Prior information plays a role in penalizing the elements of precision matrix, which may not dominate the results since it is used to restrict the search space to get more possible existing associations. If the prior information mined is rare, then the method may be considered a graphical lasso problem. However, it has not been tested on a dynamic model of microbial communities. Extending MPLasso to estimating associations among communities may, therefore, require a new algorithm.

\subsection{CD-Trace} 
D-trace loss \citep{Dtrace2014} is a convex and smooth loss function with a unique minimizer at the inverse covariance matrix. The sparse precision matrix can then be estimated by minimizing the lasso penalized D-trace loss under a positive-definite constraint. D-trace loss is proposed since graphical lasso is not computationally efficient. It can be viewed as an analogue of the least squares loss used in regression to estimate the precision matrix.

Based on D-trace loss, \citep{CDTrace2019} introduces an empirical loss function for compositional data called compositional D-trace (CD-trace) loss. Under the positive-definite constraint, minimizing lasso-penalized CD-trace loss can get a sparse matrix estimator for the direct microbial interaction network.

First, \citep{CDTrace2019} assumes that $ \log \boldsymbol{z} \sim \mathcal{N}(\boldsymbol{\mu},\Sigma) $. If the covariance matrix $\Sigma$ and the clr transformation matrix $F$ are exchangeable, then the exchangeable condition and its equivalent form can be written as:
\begin{equation}
\begin{gathered}
\Sigma F=F\Sigma \iff \\
\sum_{l}\text{cov}(\log z_{i},\log z_{l})=\sum_{l}\text{cov}(\log z_{j},\log z_{l}), \forall i,j.
\end{gathered}
\label{eq:CDTrace_exchangeEq}
\end{equation}
Hence, \citep{CDTrace2019} assumes that $ \text{var}(\log z_{i})=\text{var}(\log z_{j}), \sum_{l\neq i}\text{cov}(\log z_{i},$
$\log z_{l})\ =\sum_{l\neq j}\text{cov}(\log z_{j},\log z_{l}),\ \forall i,j=1,\dots,p.$ 
In fact, for each species, the second assumption implies that the average correlation with other species is nearly the same. This is weaker than the assumption made in SparCC \citep{SparCC2012} which requires the average correlation to be small.

Since the precision matrix $ \Theta$ can characterize the direct microbial interaction network, \citep{CDTrace2019} aims to estimate $ \Theta $ using relative abundance $ X $. Similar to the D-trace loss proposed by \citep{Dtrace2014}, \citep{CDTrace2019} proposes a convex loss function by which the minimizer is achieved at $ \Theta=\Sigma^{-1} $. Because $ \Sigma $ can not be directly estimated, \citep{CDTrace2019} uses the bridge in Equation~\ref{eq:Notations_varBridge} and replaces $ F\Sigma F $ with $ F\widehat{\Omega} F $ where $ \widehat{\Omega} $ is the finite sample estimate of $ \Omega $. Assuming that the network is sparse by adding $\ell_1$ penalty, $\widehat{\Theta}$ is obtained by the optimization problem of lasso penalized CD-trace loss:
\begin{equation}
\mathop{\min}_{\Theta\succ 0}\frac{1}{4}\left(\left\langle F\Theta,\Theta F\widehat{\Omega} F \right\rangle + \left\langle F\widehat{\Omega} F\Theta,\Theta F \right\rangle\right)- \left\langle \Theta,F \right\rangle+\lambda \left\|\Theta\right\|_{1}. 
\label{eq:CDTrace_rawObj}
\end{equation} 

The loss function proposed in \citep{CDTrace2019} is convex which makes solution easier than that in gCoda \citep{gCoda2017}. During application, \citep{CDTrace2019} adds a pseudo-count to avoid zero counts, which may result in inflated composition, and the Gaussian distribution assumption may be oversimplified. This method still needs further study in contending with excessive zeros and information from these zeros. Besides, theoretical support for the consistency of the estimator is absent. Finally, the computational complexity of CD-trace is $ O(p^{3}) $, the same as that for gCoda and SPIEC-EASI \citep{SPIECEASI2015}, which may be heavy when $ p $ is too large.

\subsection{CDTr}
Similar to CD-trace, CDTr \citep{CDTr2019} also assumes that the log-transformed absolute abundance follows multivariate normal distribution. To estimate the precision matrix $ \Theta $, a new optimization problem is proposed by incorporating clr transformation and bridge in Equation~\ref{eq:Notations_varBridge} based on D-trace \citep{Dtrace2014}, and CDTr also replaces $ F\Sigma F $ with the empirical estimator $ F\widehat{\Omega} F $, as well as adds $ \ell_{1} $-penalty on $\Theta$.
\begin{equation}
\widehat{\Theta}_{\text{CDTr}}=\mathop{\arg\min}_{\Theta\succ 0,\Theta=\Theta^{T}}\frac{1}{2}\left\langle \Theta^{2},F\widehat{\Omega} F \right\rangle-\left\langle \Theta,F \right\rangle+\lambda\left\|\Theta \right\|_{1,\text{off}}. 
\label{eq:CDTr_optimizeEq}
\end{equation}
To ensure that $ \Theta=\Sigma^{-1} $ is the minimizer of Equation~\ref{eq:CDTr_optimizeEq}, the exchange condition in Equation~\ref{eq:CDTrace_exchangeEq} should hold, although \citep{CDTr2019} states that CDTr still performs well, even when it doesn't hold.

It is worth noting that the form of Equation~\ref{eq:CDTr_optimizeEq} is simpler than that of CD-trace \citep{CDTrace2019}. To solve the problem in Equation~\ref{eq:CDTr_optimizeEq}, \citep{CDTr2019} utilizes the algorithm proposed in \citep{Dtrace2014}.

The proposed loss function has a unique solution at $ \Theta=\Sigma^{-1} $ under the exchangeable condition. Whether such exchangeable condition is reasonable needs to be further examined theoretically and biologically.  In addition, similar to SPIEC-EASI \citep{SPIECEASI2015}, gCoda \citep{gCoda2017} and CD-trace \citep{CDTrace2019}, no theoretical guarantee ensures the consistency of the estimators. 

\subsection{MInt}
\citep{MInt2016} proposes a Poisson-multivariate normal hierarchical model called MInt to capture the conditional correlation structure, in which the Poisson layer controls the confounding predictors and the multivariate normal layer with $ \ell_{1} $ regularization depicts taxon–taxon interactions. 

Let $ \widetilde{T} $ be the $ n\times o $ design matrix, in which $ \tilde{t}_{ij} $ is the value of predictor $ j $ in sample $ i $, \citep{MInt2016} assumes that: 
\begin{gather}
y_{ij} \sim \text{Poisson}\left(\exp\left\lbrace \tilde{\boldsymbol{t}}_{i\cdot}\boldsymbol{\beta}_{\cdot j}+\epsilon_{ij}\right\rbrace \right), \\
\boldsymbol{\epsilon}_{i\cdot} \sim \text{Multivarivate Normal}\left(\boldsymbol{0},\Theta \right), 
\end{gather}
where $ E=(\epsilon_{ij})_{n\times p} $ is the latent abundance matrix, and $ \boldsymbol{\beta}=(\beta_{ij})_{o\times p} $ in which $ \beta_{ij} $ is the $ i $-th predictor's coefficient for taxon $ j $. $ \boldsymbol{0} $ is a $ 1\times p $ zero vector. Then the log posterior can be written as:	
\begin{equation}
\begin{gathered}
\sum_{j=1}^{p}\sum_{i=1}^{n}\left[y_{ij}\left(\tilde{\boldsymbol{t}}_{i\cdot}\boldsymbol{\beta}_{\cdot j}+\epsilon_{ij} \right) - \exp\left\lbrace \tilde{\boldsymbol{t}}_{i\cdot}\boldsymbol{\beta}_{\cdot j}+\epsilon_{ij}\right\rbrace \right] \\
+ \frac{n}{2}\log\det \Theta-\frac{n}{2}\text{tr}\left(S(E)\Theta \right), 
\end{gathered}
\end{equation}	
in which $ S(E)=\frac{E^{T}E}{n} $ is the empirical covariance matrix of $ E $.

\citep{MInt2016} assumes that $ \mathbb{E}(\epsilon_{ij})=0, \ \forall i,j $ and that $ \widetilde{T} $ contains all relevant confounding covariates. Hence, the columns of $ E $ can be viewed as adjusted "residual" abundance measurements of the corresponding taxon after confounding predictors in $ \widetilde{T} $ are controlled. The only information that remains in these residuals must come from interactions between taxa. To model such direct interactions, \citep{MInt2016} turns to its equivalent form that models the conditional independences at the level of latent abundances. 

Last, $ \ell_{1} $-penalty is added to $ \Theta $ to make it sparse and the optimization problem, which is given in Equation~\ref{eq:PoiMultiNorm_max}, can be solved through an iterative conditional mode algorithm \citep{MInt2016} as:
\begin{equation}
\begin{gathered}
\max_{\beta,\epsilon,\Theta}\sum_{j=1}^{p}\sum_{i=1}^{n}\left[y_{ij}\left(\tilde{\boldsymbol{t}}_{i\cdot}\boldsymbol{\beta}_{\cdot j}+\epsilon_{ij} \right) - \exp\left\lbrace \tilde{\boldsymbol{t}}_{i\cdot}\boldsymbol{\beta}_{\cdot j}+\epsilon_{ij}\right\rbrace \right] \\
+ \frac{n}{2}\log\left(\det \Theta\right) -\frac{n}{2}\text{tr}\left(S(E)\Theta \right)- \frac{n\lambda}{2}\left\|\Theta \right\|_{1}+p^{2}\log\left(\frac{n\lambda}{4} \right) .
\label{eq:PoiMultiNorm_max}
\end{gathered}
\end{equation}

MInt is based on a Poisson-multivariate normal model which models the Poisson mean as a multivariate normal random variable, leading to the flexible mean-variance structure. It can also readily handle overdispersion. When inferring direct interactions, it takes controlling for covariates into consideration. However, these potential confounding variables must be diligently measured; otherwise, indirect interactions may be regarded as direct interactions. Besides, MInt directly handles count data without considering the compositional nature of data.

\subsection{mLDM}
\citep{mLDM2017} proposes a metagenomic Lognormal-Dirichlet-Multinomial hierarchical model (mLDM) to capture the conditional dependency between two taxa, and the association between taxon and environment factors. 

mLDM assumes that each piece of count data $\boldsymbol{y}_{i \cdot}$ follows a multinomial distribution. It considers that absolute abundances are influenced by environmental factors and associations among microbes. The generative process of the hierarchical mLDM model can be concluded as follows:
\begin{equation}
\begin{gathered}
\boldsymbol{y}_{i \cdot} \sim \text{Multinomial}(\boldsymbol{q}_{i \cdot}), \\
\boldsymbol{q}_{i \cdot} \sim \text{Dirichlet}(\boldsymbol{z}_{i \cdot}), \\
\boldsymbol{z}_{i \cdot}=\exp\left(B^{T}\boldsymbol{m}_{i\cdot}+\boldsymbol{v}_{i\cdot} \right), \\
\boldsymbol{v}_{i\cdot} \sim \mathcal{N}(B_{0},\Sigma),
\end{gathered}
\end{equation}
where $ \boldsymbol{q}_{i \cdot} $ is the multinomial parameter, $ \boldsymbol{z}_{i \cdot} $ is the absolute abundance for all taxa in the $ i $-th sample, $ \boldsymbol{m}_{i\cdot} $ is the environmental factor, and $B_0$ is a $p$-dimensional vector.
Let $ V=(v_{ij})_{n\times p} $, of which the $ i $-th row is $ \boldsymbol{v}_{i\cdot} $. Then the posterior of $ V $ can be written as:
\begin{equation}
\begin{gathered}
p(V|Y,M,B,B_{0},\Theta)\propto \\\prod_{i=1}^n p(\boldsymbol{y}_{i \cdot}|\boldsymbol{z}_{i \cdot})\cdot p(\boldsymbol{z}_{i \cdot}|\boldsymbol{v}_{i \cdot},B,B_0,\boldsymbol{m}_{i\cdot})\cdot p(\boldsymbol{v}_{i \cdot}|B_0,\Theta) \\  
\end{gathered} 	
\end{equation}

Because of the multivariate normal distribution assumption, the precision matrix $ \Theta=(\theta_{ij})_{p\times p} $ can characterize the conditional dependence associations among taxa. Generally, interactions among taxa are not dense, and only a small number of environmental factors can assume a predominant role in affecting taxa. Hence, \citep{mLDM2017} assumes that graphs are sparse \textit{via} $ \ell_{1} $ penalty. The final optimization problem is:
\begin{equation}
\min_{B,B_{0},\Theta,V} -\frac{1}{n}\log p(V|Y,M,B,B_{0},\Theta)+\frac{\lambda_{1}}{2}\left\|\Theta \right\|_{1}+ \lambda_{2}\left\|B \right\|_{1}.
\end{equation}

mLDM estimates both the conditional associations among microbes and the associations between microbes and environmental factors. It considers compositional bias and variance of the metagenomic data. mLDM assumes that microbes have linear relationships with environmental factors, which may not, in fact, be the case even though \citep{mLDM2017} argues that mLDM can, to some extent, capture some nonlinear associations to some extent. Besides, many parameters are introduced in mLDM, that can limit its efficiency and scalability. The complexity of mLDM also reduces its interpretability. Moreover, \citep{mLDM2017} lacks theoretical justification of the estimators.

\subsection{HARMONIES}
Considering that microbiome sequencing data usually have different sequence depths across samples, many zeros, and over-dispersion, normalization is an important step before analysis. Unlike traditional normalization which gets the compositional data first and then applies log-ratio transformation \citep{SPIECEASI2015}, \citep{HARMONIES2020} proposes a model-based normalization method designed for microbiome count data called HARMONIES. 

The normalization step of HARMONIES is realized by building a zero-inflated negative binomial (ZINB) model with Dirichlet process prior (DPP). It can capture the effect of sequencing depth and distinguish two types of zero count. One is true zeros and the other is caused by failure to detect. A multiplicative characterization is applied to the mean of negative binomial (NB), which decomposes the mean into a sample-specific size factor and the normalized relative abundance. This is performed to justify the latent heterogeneity in microbiome data. Then HARMONIES adopts the Dirichlet process prior (DPP) to estimate the sample-specific size factor to capture variations in sequencing depth across samples. Thus, sample heterogeneity can be taken account for by DPP. After using Markov chain Monte Carlo (MCMC) to estimate model parameters, the normalized relative abundance after denoising $ A=(\alpha_{ij})_{n\times p} $ is obtained. 

\citep{HARMONIES2020} assumes that $ \log\boldsymbol{\alpha}_{i\cdot},i=1,\dots,n $ follows a multivariate normal distribution $ \mathcal{N}(\boldsymbol{\mu},\Sigma) $ independently. Then HARMONIES uses graphical Lasso to estimate the precision matrix $ \Theta=\Sigma^{-1} $ similar to SPIEC-EASI \citep{SPIECEASI2015} \textit{via} the optimization problem:
\begin{equation}
\mathop{\min}_{\Theta\succ 0}-\log\left(\det\Theta \right)+\text{tr}\left(S\Theta \right)+\lambda\left\|\Theta\right\|_{1},
\end{equation} 
where $ S $ is the sample covariance matrix based on $ \log A $.  

HARMONIES takes into account excessive zeros and over-dispersion by modeling the count data with a zero-inflated negative binomial distribution. It also takes sample heterogeneity into account by further implementing DPPs. However, in some cases, the sample size obtained is always small, and this limitation has a direct effect on the estimated normalized matrix $ A $ from the ZINB model. Specifically, for a taxon $i$, a small sample size could lead to a large variance in the posterior distribution of $\log\boldsymbol{\alpha}_{i\cdot}$.

\subsection{BC-GLASSO}
To tackle the problem of different sequencing depths, a widely used practice is utilizing compositional proportions calculated by count data as a proxy for the true relative abundance, such as SPIEC-EASI \citep{SPIECEASI2015}. However, doing so ignores the variation in microbial counts. In order to reduce bias in the estimator coming from the proportion-based approach, \citep{BCGLASSO2020} models the count data directly to quantify the bias and proposes bias-corrected graphical lasso (BC-GLASSO).

BC-GLASSO has two steps and is similar to SPIEC-EASI \citep{SPIECEASI2015}. First, BC-GLASSO assumes that the count data $ \boldsymbol{y}_{i\cdot} \sim \text{Multinomial}(\boldsymbol{\alpha}_{i\cdot})$, where $ \sum_{j=1}^{p}\boldsymbol{\alpha}_{ij}=1 $. Then, \citep{BCGLASSO2020} applies alr transformation \citep{coda1986} to $ \boldsymbol{\alpha}_{ij},i=1,\dots,n$ as shown in Equation~\ref{eq:Notation_alr}, and \citep{BCGLASSO2020} further assumes that the alr transformed $\boldsymbol{h}_{i\cdot}\sim \mathcal{N}(\boldsymbol{\mu},\Sigma)$. Using alr transformation can make the precision matrix $ \Theta $ positive-definite, while clr transformation may lead to a singular covariance matrix \citep{SPIECEASI2015}. 

The next step of BC-GLASSO is to estimate $ \Theta $. At the beginning of estimation, \citep{BCGLASSO2020} applies the same alr transformation to the count data as the estimation of $ \boldsymbol{h}_{i\cdot} $, satisfying $\hat{h}_{ij}=\log\left(\frac{\alpha{ij}}{\alpha_{ip}}\right)$. However, as an estimator of $ \boldsymbol{h}_{i\cdot}$, $ \hat{\boldsymbol{h}}_{i\cdot}$ may induce variation. This prompts BC-GLASSO to develop a bias correction procedure for estimated covariance.

Let $ \bar{h}_{k}=\frac{1}{n}\sum_{i=1}^{n}\hat{h}_{ik}\ ,\hat{\sigma}_{i,k,l}=\left(\hat{h}_{ik}-\bar{h}_{k}\right)\left(\hat{h}_{il}-\bar{h}_{l}\right)$. Then $ \hat{\sigma}_{kl}=\frac{1}{n}\sum_{i=1}^{n}\hat{\sigma}_{i,k,l} $. As the estimator of $ \sigma_{i,k,l} $, \citep{BCGLASSO2020} points out that $ \hat{\sigma}_{i,k,l} $ has an approximate bias with the order of $ s(\boldsymbol{y}_{i\cdot})^{-1} $. 
Inspired by the approximation equation of $ \text{E}(\hat{\sigma}_{i,k,l}) $, \citep{BCGLASSO2020} fits a linear regression to correct bias and uses the least squares estimator as the estimator of $ \hat{\sigma}_{kl} $, which is denoted as $ \tilde{\sigma}_{kl} $. In this way, the bias-corrected estimator $ \tilde{\sigma}_{kl} $ is the weighted mean of $ \hat{\sigma}_{i,k,l} $. From the illustration in \citep{BCGLASSO2020}, $ \tilde{\sigma}_{kl} $ is also approximately unbiased when all sequencing depths and sample sizes are large. Hence, BC-GLASSO uses the covariance matrix $ \widetilde{\Sigma}=(\tilde{\sigma}_{kl})_{p\times p} $ to replace $ \widehat{\Sigma} =(\hat{\sigma}_{ij})_{p\times p} $, which is the sample covariance of $ \hat{\boldsymbol{h}}_{i\cdot},i=1,\dots,n $, and the sparse estimator of $ \Theta $ can be achieved by solving:
 \begin{equation}
 	\min_{\Theta} -\log\left(\det\Theta \right)+\text{tr}\left(\widetilde{\Sigma}\Theta\right)+\lambda\left\|\Theta\right\|_{1}.
 \end{equation}

The covariance matrix in BC-GLASSO is defined on the alr transformed multinomial probabilities instead of clr transformed ones. This transformation has the advantage of leading to a positive definite covariance matrix and thus makes the precision matrix well-defined. BC-GLASSO is, in fact, a hierarchical model. The multinomial distribution can capture technical variation of the count data, and the logistic normal distribution can detect biological variation of microbiome composition. BC-GLASSO considers the bias induced by the estimator and correct such bias. However, BC-GLASSO does not now take covariates, such as sex and age, into consideration. Since alr transformation needs a reference taxon, the potential interactions between the reference taxon and other taxa can not be estimated and analyzed using BC-GLASSO. 

\subsection{CompGlasso and its reference invariant version}
The compositional graphical lasso (CompGlasso) proposed in \citep{CompGlasso2020} is also based on Logistic Normal Multinomial distribution such as that found in BC-GLASSO \citep{BCGLASSO2020}. The invariance of the proposed method with respect to the choice of reference taxon for the alr transformation is further explored in \citep{InvGlasso2020}, which has not been completely studied before.
    
First, \citep{CompGlasso2020} proposed compositional graphical lasso based on the logistic normal multinomial model. A sparse estimator of $\Omega$ is found by maximizing the maximum posterior distribution with $\ell_{1}$-penalty, which equals minimizing the following objective function: 
    \begin{equation}
    	\begin{gathered}
    	\mathcal{L}\left( \boldsymbol{h}_{1\cdot},\dots,\boldsymbol{h}_{n\cdot},\boldsymbol{\mu},\Theta\right) \\
    	=-\frac{1}{n}\sum_{i=1}^{n}\left[\boldsymbol{y}_{i,-(p)}^{T}\boldsymbol{h}_{i\cdot}-s(\boldsymbol{y}_{i\cdot})
        \log\left(\boldsymbol{1}_{p-1}^{T}\exp(\boldsymbol{h}_{i\cdot})+1 \right) \right]  \\
    	+\frac{1}{2n}\sum_{i=1}^{n}\left(\boldsymbol{h}_{i\cdot}-\boldsymbol{\mu}\right)^{T}\Theta\left(\boldsymbol{h}_{i\cdot}-\boldsymbol{\mu}\right)-\frac{1}{2}\log\left(\det\Theta\right) +\lambda \left\|\Theta\right\|_{1},
    	\end{gathered}
    	\label{eq:CompGlasso_loss}
    \end{equation}
where $ \boldsymbol{y}_{i,-(p)}=(y_{i1},\dots,y_{i,p-1})^{T} $. The first term in Equation~\ref{eq:CompGlasso_loss} is the negative log-likelihood, and the remaining terms are the normal objective function of graphical lasso for the multivariate normal distribution considering $\boldsymbol{h}_{1\cdot},\cdots,\boldsymbol{h}_{n\cdot}$ as known quantities.

However, a different choice of reference may result in a different objective function. Either invariance or robustness of the method with regard to the choice of reference taxon is one of the major concerns about alr transformation. From \citep{Aitchison1986}, the first three terms in Equation~\ref{eq:CompGlasso_loss} remain invariant when choosing the $(p-1)$-th entry as the reference. \citep{InvGlasso2020} states that the $ (p-2)\times(p-2) $ upper-left submatrix of both the inverse covariance matrix and its estimator of the alr transformed vectors, is invariant to choosing the $ p $-th or $ p-1 $-th entry as the reference.

Generally, $ \boldsymbol{\alpha}_{\mathcal{R}} $ is denoted as "candidate references" and $ \mathcal{R} $ is its index set. $ \mathcal{R}^{C} $ is the complementary set of $ \mathcal{R} $. The reference-invariant property can be stated as follows: for any alr tranformation using a reference in the set of candidate references $ \boldsymbol{\alpha}_{\mathcal{R}} $, the $ |\mathcal{R}^{C} |\times|\mathcal{R}^{C}| $ upper-left submatrix of the (estimated) inverse covariance matrix of the alr transformed data is invariant to the choice of the reference. Therefore, by replacing the last term $ \lambda \left\|\Theta\right\|_{1} $ with $\lambda \left\|\Theta_{\mathcal{R}^{C},\mathcal{R}^{C}}\right\|_{1}$ in Equation~\ref{eq:CompGlasso_loss}, we can obtain the reference-invariant objective function $\mathcal{L}_{inv}\left( \boldsymbol{h}_{1\cdot},\dots,\boldsymbol{h}_{n\cdot},\boldsymbol{\mu},\Theta\right)$, which is defined as the reference-invariant version of compositional graphical lasso.
    
The trick played here is only penalizing the invariant part of the inversion covariance matrix. Therefore, by applying this trick to graphical lasso, \citep{Glasso2008} can get the reference-invariant graphical lasso (Inv-gLASSO) \citep{InvGlasso2020}, of which the objective function is:
    \begin{equation}
    	\begin{gathered}
    	\mathcal{L}_{inv}\left(\boldsymbol{\mu},\Theta\right) =\frac{1}{2n}\sum_{i=1}^{n}\left(\boldsymbol{h}_{i\cdot}-\boldsymbol{\mu}\right)^{T}\Theta\left(\boldsymbol{h}_{i\cdot}-\boldsymbol{\mu}\right) \\ -\frac{1}{2}\log\left(\det\Theta\right)+\lambda \left\|\Theta_{\mathcal{R}^{C},\mathcal{R}^{C}}\right\|_{1}.
    	\end{gathered}
    	\label{eq:InvGlasso_Obj}
    \end{equation}

Compositional graphical lasso only has two levels to increase its interpretability. \citep{CompGlasso2020} also establishes the theoretical convergence of the algorithm. The reference-invariant property clearly shows how different reference taxon choices affect the estimated network. Based on this property, a reference-invariant optimization problem is proposed, making the alr transformation free from the choice of the reference taxon. However, the empirical reference-invariant inverse covariance matrix and network may depend on the algorithm used to optimize the objective function.
 
\section{Mixture network}
The methods described above all assume that the sample-taxa matrix is only related to a single network topology and set of edge weights. A sample-taxa matrix, however, is constructed by various samples in which taxa may be involved in more than one biological network. Moreover, different environmental conditions can lead to different association networks. In particular, for human gut microbiome data, other variables, such as age, sex and diet, can also influence microbial network interactions, changing them according to these variables.
Therefore, Tavakoli et al. \citep{MixMPLN2019} propose a mixture model named MixMPLN which is comprised of $ K $ Multivariate Poisson Log-Normal (MPLN) distributions. The mixture model directly tackles the count data, and the precision matrix of each component represents a microbial network. MixMPLN estimates networks \textit{via} maximizing the likelihood function added with $ \ell_{1} $- penalty for sparsity. Furthermore, Tavakoli et al. \citep{MixGGMMixMCMC2019} propose two new algorithms to infer multiple microbial networks. One is MixMCMC which estimates latent parameters of the mixture MPLN model via Markov Chain Monte Carlo (MCMC) sampling. It differs from MixMPLN in that MixMPLN estimates parameters using the gradient descent method. The Poisson layer in the MPLN model may include covariates to control for confounding factors \citep{Biswas2015} and can be studied further. Another one is MixGGM which handles compositionality by adopting clr-transformation. MixGGM assumes that such transformed data follow a mixture of Multivariate Gaussian distributions. Both MixMCMC and MixGGM add $ \ell_{1} $-penalty to infer sparse networks, and their optimization algorithms are based on the MM principle. However, their computation complexity of them has not been well-analyzed. 

As mentioned before, environmental factors, such as disease, lifestyle and host genotype \citep{Qin2012}, 
may result in the variation of abundances. mLDM proposed by Yang et al. \citep{mLDM2017} removes the indirect associations caused by common environmental factor, and estimates both direct associations between two taxa and associations between taxa and environmental factors. However, this model assumes only one underlying biological network which may not hold, as previously indicated. Thus Yang et al. \citep{kLDM2020} proposed a k-Lognormal-Dirichlet-Multinomial model (kLDM) which is a hierarchical Bayesian model. This model can automatically determine the number of environmental conditions. Here the environmental condition is defined as some specific range of values of environmental factors in \citep{kLDM2020}. kLDM can estimate the conditionally dependent taxon-taxon associations and taxon-environmental factor associations under each environmental condition. For modeling, kLDM also takes compositional bias and the large variance of count data into consideration. However, kLDM shares a problem in common with mLDM in that too many parameters may lead to inefficient computation. This, in turn, results in unsatisfactory scalability of kLDM.

\subsection{MixMPLN and MixMCMC}
\subsubsection{MixMPLN}
A generated sample-taxa matrix may be part of a large project, so it may be affected by more than one microbial network owing to different experimental conditions. \citep{MixMPLN2019} proposes a mixture model that assumes the count data are all generated from a mixture of $K$ Multivariate Poisson Log-Normal (MPLN) distributions \citep{Aitchison1989}. 
It first assumes that $ \boldsymbol{y}=(y_{1},\dots,y_{p}) $ is a random count vector which follows a single MPLN distribution:
\begin{equation}
\begin{gathered}
y_j|\lambda_{j} \sim \text{Poisson}(\lambda_{j}),\ j=1,\cdots,p\\
(\lambda_{1},\dots,\lambda_{p})^{T} \sim \mathcal{N}\left(\boldsymbol{\mu},\Sigma \right). 
\end{gathered}
\end{equation}

Then, for the mixture model, the probability density function of the $ l $-th MPLN distribution is denoted as $ p_{l}(\boldsymbol{y}_{i\cdot}|\Xi_{l}) $
, in which $ \Xi_{l}=(\boldsymbol{\mu}_{(l)},\Sigma_{(l)}) $ is the parameter set of the multivariate normal distribution. Then the joint probability density function of $n$ samples is:
\begin{equation}
p\left(Y|\Pi,\Xi\right)=\prod_{i=1}^{n}\sum_{l=1}^{K}\pi_{l}p_{l}(\boldsymbol{y}_{i\cdot}|\Xi_{l}),
\end{equation}
where $ \Pi=(\pi_{1},\dots,\pi_{K})$, $\Xi=(\Xi_{1},\dots,\Xi_{K}) $.

Let $ \boldsymbol{\lambda}_{il} $ be the $ p $-dimensional latent variable associated with $ \boldsymbol{y}_{i\cdot} $ in the $ l $-th component.The joint density function is then denoted as $ p_{l}(\boldsymbol{y}_{i\cdot}|\boldsymbol{\lambda}_{il},\Xi_{l}) $, and the marginal density function is $ p_{l}(\boldsymbol{y}_{i\cdot}|\Xi_{l})=\int_{\mathbb{R}^{p}}p_{l}(\boldsymbol{y}_{i\cdot}|\boldsymbol{\lambda}_{il},\Xi_{l})d\boldsymbol{\lambda}_{il} $. Based on estimated parameters,  $\prod_{i=1}^{n}p_{l}(\boldsymbol{y}_{i\cdot}|\boldsymbol{\lambda}_{il},\Xi_{l})$ can be used to approximate $p_{l}(\boldsymbol{y}|\Xi_{l})$. Then \citep{MixMPLN2019} follows the MM principle \citep{Hunter2004}\citep{Lange2016} to maximize the objective function $ \mathcal{L} $ at the $ t+1 $-th step as follows:
\begin{equation}
\begin{aligned}
\mathcal{L}\left(\Pi,\Lambda,\Xi \right)=&-\sum_{i=1}^{n}\sum_{l=1}^{K}\omega_{il}^{t}\log\omega_{il}^{t}+\sum_{i=1}^{n}\sum_{l=1}^{K}\omega_{il}^{t}\log\pi_{l}\\
&+\sum_{i=1}^{n}\sum_{l=1}^{K}\omega_{il}^{t}\log p_{l}(\boldsymbol{y}_{i\cdot}|\lambda_{il},\Xi_{l}),
\end{aligned}     
\end{equation} 
where $ \omega_{il}^{t}=\frac{\pi_{l}^{t}p_{l}(\boldsymbol{y}_{i\cdot}|\boldsymbol{\lambda}_{il}^{t},\Xi_{l}^{t})}{\sum_{l=1}^{K}\pi_{l}^{t}p_{l}(\boldsymbol{y}_{i\cdot}|\boldsymbol{\lambda}_{il}^{t},\Xi_{l}^{t})} $. 
\citep{MixMPLN2019} also extends the MixMPLN framework by adding sparse constraint onto $K$ precision matrices and during iteration, graphical lasso \citep{Glasso2008} is applied to each component of the mixture model to estimate the precision matrix.

MPLN distribution which is the basis of MixMPLN, can potentially include technical and biological covariates in the Poisson layer to control for confounding factors. However, as a mixture model, MixMPLN contains many parameters which can increase the time of optimization.

\subsubsection{MixMCMC}
MixMCMC \citep{MixGGMMixMCMC2019} also models the sample-taxa count data $ Y $ by a mixture of MPLN distributions and follows the MM principle to simplify the objective function. It differs from MixMPLN in estimating the latent variables. MixMPLN estimates all parameters ($\Xi,\Lambda$) \textit{via} gradient descent, while MixMCMC utilizes gradient descent to estimate $ \Xi $, along with Markov Chain Monte Carlo sampling, to estimate the latent variables $ \Lambda $.

MixMPLN could identify both positive and negative interactions between pairs of taxa for selected threshold values. However, the biological significance of these interactions should be investigated further.
MixMPLN tackles the count data directly which considers different microbial clusters. Technical covariates, or environmental factors, which may influence a microbe network can be included in the MixMPLN structure. Nonetheless, the Bayesian approaches to select the model need further study, and the MCMC process in MixMCMC may be time-consuming. 

\subsection{MixGGM}
\citep{MixGGMMixMCMC2019} proposes MixGGM which applies a log-ratio transformation to the sample count matrix and then models the problem
as a mixture of Multivariate Gaussian distributions. That is to say, the density function of the clr transformed data $ W $ is:
\begin{equation}
p\left(W|\Pi,\Xi\right)=\prod_{i=1}^{n}\sum_{l=1}^{K}\pi_{l}p_{l}(\boldsymbol{w}_{i\cdot}|\Xi_{l}). 
\end{equation}
Here $ p_{l}(\boldsymbol{w}_{i\cdot}|\Xi_{l}) $ is the density of $ \mathcal{N}(\boldsymbol{\mu}_{(l)},\Sigma_{(l)}) $ and $ \Xi_{l}=(\boldsymbol{\mu}_{(l)},\Sigma_{(l)}) $.

Based on the MM principle \citep{Hunter2004}\citep{Lange2016}, MixGGM maximizes the function as follows:
\begin{equation}
\mathcal{L}\left(\Pi,\Xi \right)=\sum_{i=1}^{n}\sum_{l=1}^{K}\omega_{il}^{t}\log\left(\frac{\pi_{l}}{\omega_{il}^{t}}p_{l}(\boldsymbol{w}_{i\cdot}|\lambda_{il},\Xi_{l}) \right), 
\end{equation}
where $ \omega_{il}^{t}=\frac{\pi_{l}^{t}p_{l}(\boldsymbol{w}_{i\cdot}|\boldsymbol{\lambda}_{il}^{t},\Xi_{l}^{t})}{\sum_{l=1}^{K}\pi_{l}^{t}p_{l}(\boldsymbol{w}_{i\cdot}|\boldsymbol{\lambda}_{il}^{t},\Xi_{l}^{t})} $. The iterative process of estimating parameters is similar to that in MixMPLN \citep{MixMPLN2019}.

\citep{MixGGMMixMCMC2019} also extends MixGGM by adding sparse assumption onto the precision matrix for each multivariate Gaussian distribution. This is realized by applying graphical lasso \citep{Glasso2008} to all $ K $ components. 

MixGGM is based on the idea that covariance matrix of clr transformed data approximates that of log-transformed absolute abundances. So, it may share the same problem as SPIEC-EASI \citep{SPIECEASI2015} since approximation strongly depends on the conditional number of the inverse covariance matrix. Besides, the clr transformed data may not fitting sub-Gaussian or Gaussian distribution well. 

\subsection{kLDM}
Since the distribution of environmental conditions may be complex and continuous, the k-Lognormal-Dirichlet-Multinomial (kLDM) model proposed by \citep{kLDM2020} estimates multiple association networks with respect to specific environmental conditions.

kLDM assumes that environmental factors regulate associations among microbes. These environmental factors tend to be stable given identical environmental conditions, but vary with environmental conditions. It approximates the association pattern with mixtures of multiple clusters, where each cluster represents an environmental condition. The hierarchical structure of kLDM is:
\begin{equation}
\begin{gathered}
\boldsymbol{y}_{i\cdot}\sim \text{Multinomial}(\boldsymbol{\alpha}_{i\cdot}),\\
\boldsymbol{\alpha}_{i\cdot}\sim \text{Dirichlet}(\boldsymbol{z}_{i\cdot}),\\
\boldsymbol{z}_{i\cdot}|c_{i}=\exp\left(B^{(c_{i})T}\boldsymbol{m}_{i\cdot}+\boldsymbol{v}_{i\cdot}\right),\\
\boldsymbol{v}_{i\cdot}|c_{i}\sim \mathcal{N}(B_{0}^{(c_{i})},(\Theta^{(c_{i})})^{-1}),\\
\boldsymbol{m}_{i\cdot}|c_{i}\sim \mathcal{N}(\boldsymbol{\nu}^{(c_{i})},\Gamma^{(c_{i})}),\\ 
c_{i}\sim\text{Categorial}(\pi_{1},\pi_{2},..., \pi_{K}),
\end{gathered}
\end{equation}
where parameter $ \boldsymbol{\alpha}_{i\cdot} $  represents the true underlying relative abundance, $ \boldsymbol{z}_{i\cdot} $ represents the underlying absolute abundance, $ c_{i} $ represents a certain environmental condition, while  $ \boldsymbol{m}_{i\cdot} $ is the $ i $-th environment factor.

The direct associations between microbes under the $ k $-th environmental condition are characterized by $ \Theta^{(k)} $, and 
$ B^{(k)} $ represents the associations between microbes and $k$-th environmental factors. \citep{kLDM2020} further assumes that these association networks are sparse. Then it proposes a split-and-merge algorithm to estimate the number of environmental conditions and association networks.

Similar to mLDM, kLDM estimates both associations among microbes and associations between microbes and environmental factors. It can determine the number of environmental conditions automatically. However, the mathematical score, such as the EBIC score used in \citep{kLDM2020} to merge clusters, may not be sensitive enough to separate two clusters that represent two environmental conditions. kLDM assumes that environmental factors follow a Gaussian distribution, but this assumption may not hold given categorical data. Similar to mLDM \citep{mLDM2017}, kLDM contains too many parameters. The complexity may affect its scalability and efficiency, as well as interpretability. It can be further improved by incorporating prior information and gaining the ability of including rarer OTU counts with low occurrence.

\section{Differential network}
Mixture network inference methods estimate multiple underlying microbial networks simultaneously. In addition to considering the general case that multiple networks are associated with the sample-taxa matrix, discovering how the interaction network varies with environmental conditions or other confounding factors is another study of much significance to researchers. The differential network is often defined as the difference between two precision matrices. However, using the difference between these two estimators of precision matrix as the estimated differential network can be problematic. One reason is that the estimator of a single precision matrix is usually obtained under the sparsity assumption which may not be true in some cases. Directly minusing such two estimators may lead to false positive and negative interactions \citep{DiffDtrace2017}. Even if the sparsity assumption were to be held, these methods would have most power in exploring strong interactions in a single condition. Thus, when detecting interactions that are not strong in one condition, but change strongly in other conditions, the power of these methods is limited \citep{DiffDtrace2017}. 

Several studies have estimated differential networks for absolute abundance data. One such method of estimating differential networks shrinks the precision matrices, as well as their difference, to perform simultaneous estimation, such as that reported in \citep{Guo2011} \citep{Chiquet2011} \citep{Danaher2014}. However, they may not perform well when precision matrices are not sparse. 
Another method estimates differential networks without sparsity assumption on each single precision matrix, such as the new $ \ell_{1} $-minimization method proposed in \citep{Zhao2014}. However, such methods have a high requirement of memory and are not computationally efficient. Therefore, Yuan et al. \citep{DiffDtrace2017} generalize the D-trace loss \citep{Dtrace2014} into a new convex and smooth loss function to estimate the difference of two precision matrices directly. Such estimation can be conducted for high-dimensional data by adding lasso penalty. A convergency theory for polynomial-tailed and sub-Gaussian distributions under a less strict irrepresentability condition than that in \citep{Dtrace2014} is also proposed by Yuan et al. \citep{DiffDtrace2017}. The new D-trace loss function in \citep{DiffDtrace2017} can be generalized to compare more than two precision matrices. However, it cannot manage situations in which precision matrices depend on continuous variables. 

Since microbiome data only carry relative abundance information, absolute abundance data is unobserved. So, methods need to be modified to adapt to compositional data. As mentioned before, He and Deng \citep{CDTr2019} propose CDTr by incorporating clr transformation into D-trace \citep{Dtrace2014} to derive the conditional correlation network. They further generalize their method and propose DCDTr by combining clr transformation with the new D-trace loss in \citep{DiffDtrace2017} to estimate the differential network for compositional data. Like CDTr, the exchangeable condition under which the DCDTr loss works has not been examined in biology. Also, no theory thus far guarantees the consistency of the DCDTr estimator. 

Microbiome Differential Network Estimation (MDiNE) \citep{MDiNE2020} is a hierarchical model which can estimate network changes with regard to a binary covariate defining groups. It assumes that the count data follow a multinomial distribution, the probabilities of which are determined by Gaussian random variables. Sparse precision matrices over these latent variables represent co-occurrence networks. The model is estimated through Hamiltonian Monte Carlo (HMC). In the field of network inference, underestimation of true associations may occur when the sample size is relatively small. However, the number of parameters of MDiNE would increase rapidly as the number of samples increases, bringing in heavy computational burden. Furthermore, MDiNE considers the estimation of the differential network of two networks using a binary covariate, which can be extended to explore how networks vary with continuous variables.

Cai et al. \citep{MarRanField2019} turn to the use Markov random field theory to overcome problems of Gaussian graphical modeling mentioned in the section on conditional correlation networks. They first discretize each composition vector into a binary one. Then, conditional dependencies between components in the vector are modeled by a binary Markov random field. To test whether two precision matrices are identical, Cai et al. \citep{MarRanField2019} conduct a global test first. If the null hypothesis is rejected, then the multiple tests of each entry are performed. The binary Markov random field used in \citep{MarRanField2019} accounts for excessive zeros in the compositional data and the unit-sum constraint. The model proposed does not require the assumption of data distribution, which, in turn, increases its robustness. Besides, tests proposed by Cai et al. \citep{MarRanField2019} can assign statistical significance to each element of the differential network. Nevertheless, the method is strongly affected by the discretization threshold chosen, and each element of the discretized vector can only take two values. 

\subsection{DCDTr}
As noted in the conditional correlation network section, \citep{CDTr2019} proposes a CDTr loss function by incorporating clr transformation into D-trace in \citep{Dtrace2014} to infer direct interactions for compositional data. Furthermore, to infer microbiome differential networks, while avoiding the use of unobserved absolute abundance data, \citep{CDTr2019} proposes a loss function called DCDTr loss by combining clr transformation with new D-trace in \citep{DiffDtrace2017}. The DCDTr loss proposed is smooth and convex and can estimate the precision matrix difference directly with compositional data instead of estimating the precision matrices individually.

Denote $ \Sigma^{(i)} $ as the covariance matrix of group $ i $, $ i=1,2 $. Then the differential network can be characterized by the difference matrix $ \Delta=(\Sigma^{(2)})^{-1}-(\Sigma^{(1)})^{-1}=\Theta^{(2)}-\Theta^{(1)} $. After being given two necessary conditions, the minimizer of this loss function is achieved at $ \Delta$. Suppose that $ \widehat{\Sigma}^{(i)} $ is the sample version of $ \Sigma^{(i)}$. To account for high-dimensional data, \citep{DiffDtrace2017} adds the $ \ell_{1} $ penalty to the difference matrix, and the extended D-trace-based optimization objective proposed in \citep{DiffDtrace2017} is:
    \begin{equation}
    \begin{gathered}
    	\min_{\Delta}\frac{1}{4}\left(\left\langle \widehat{\Sigma}^{(1)}\Delta,\Delta\widehat{\Sigma}^{(2)}\right\rangle+ \left\langle \widehat{\Sigma}^{(2)}\Delta,\Delta\widehat{\Sigma}^{(1)}\right\rangle \right) \\
    	-\left\langle\Delta,\widehat{\Sigma}^{(1)}-\widehat{\Sigma}^{(2)} \right\rangle+\lambda \left\|\Delta\right\|_{1} . 
    	\label{eq:DiffDtrace_RawOpt}
    \end{gathered}
    \end{equation}
Here $ \lambda \geq 0 $ is the tuning parameter.

To merge the compositional data transformations into D-trace loss for differential network estimation \citep{DiffDtrace2017} mentioned above, \citep{CDTr2019} first assumes that $ \log \boldsymbol{z}_{k \cdot}^{(i)}\sim N(\boldsymbol{\mu}^{(i)},\Sigma^{(i)}),k=1,\dots,n $, $ i=1,2 $. Thus, the resultant differential network is $ \Delta=(\Sigma^{(2)})^{-1}-(\Sigma^{(1)})^{-1}=\Theta^{(2)}-\Theta^{(1)} $.
  
Since true absolute abundance cannot be achieved, \citep{CDTr2019} utilizes the relationship formula between absolute and relative abundance, namely Equation~\ref{eq:Notations_varBridge}. Following Equation~\ref{eq:DiffDtrace_RawOpt}, \citep{CDTr2019} replaces $ \widehat{\Sigma}^{(i)}$  with $ F\widehat{\Omega}^{(i)}F $, $ i=1,2 $. 
Similar to CDTr, $\Delta$ is the minimizer of DCDTr under the exchangeable condition: $F\Sigma^{(i)}=\Sigma^{(i)}F,i=1,2$. \citep{CDTr2019} also assumes that the difference matrix is sparse and uses $ \ell_{1} $ penalty on it rather than requiring sparsity for each precision matrix. Therefore, the optimization problem is:
    \begin{equation}
    	\begin{gathered}
    	\min_{\Delta=\Delta^{T}} \frac{1}{4}\left(\left\langle F\widehat{\Omega}^{(1)}F\Delta,\Delta F\widehat{\Omega}^{(2)}F\right\rangle + \left\langle F\widehat{\Omega}^{(2)}F\Delta,\right. \right. \\
    	\left. \left. \Delta F\widehat{\Omega}^{(1)}F\right\rangle \right) 
    	-\left\langle\Delta,F\left(\widehat{\Omega}^{(1)}-\widehat{\Omega}^{(2)}\right)F \right\rangle
    	+ \lambda \left\|\Delta \right\|_{1}. 
    	\end{gathered}
    	\label{eq:DCDTr_opt}
    \end{equation}

DCDTr utilizes the $ F $ matrix in the bridge equation~\ref{eq:Notations_varBridge} to apply D-trace in \citep{DiffDtrace2017} onto compositional data. Similar to CDTr, the loss function proposed in DCDTr has a unique minimizer at $ \Theta=\Sigma^{-1} $ if the exchangeable condition is satisfied. However, the reasonableness of the exchangeable condition still needs to be further studied. Most importantly, no theoretical guarantee on the consistency of the estimator is provided. Besides, since DCDTr is based on the generalized D-trace loss in \citep{DiffDtrace2017}, they are equally disadvantaged in that they can neither judge whether precision matrices depend on continuous variables nor can they infer such dependencies when precision matrices depend on continuous variables.

\subsection{MDiNE}
\citep{MDiNE2020} builds a Bayesian hierarchical model called Microbiome Differential Network Estimation (MDiNE), which can estimate the co-occurrence network of two groups simultaneously with respect to a binary covariate and thus estimate the difference between them. \citep{MDiNE2020} models the counts of individual taxa through a multinomial distribution with its probabilities depending on a latent Gaussian random variable. In addition, the co-occurrence network among taxa is determined by a sparse precision matrix over all the latent items.
    
Suppose two groups have a total number of samples $ N=n_{1}+n_{2} $. Here, $ c_{i}=1 $ indicates that $ \boldsymbol{y}_{i\cdot} $ belongs to the second group, otherwise $ c_{i}=0 $. $\widehat{\boldsymbol{h}}_{i\cdot}$ is the alr transformed probabilities of the multinomial distribution, choosing the last taxon as the reference. Let $ \Theta^{(k)}=(\theta_{jl}^{(k)})_{p\times p}=(\Sigma^{(k)})^{-1} $. Then the overall model of MDiNE is:
     \begin{equation}
     	\begin{gathered}
     	\boldsymbol{y}_{i\cdot}|\boldsymbol{\alpha}_{i\cdot},\boldsymbol{h}_{i\cdot},B,\Theta^{(1)},\Theta^{(2)},\lambda\sim \text{Multinomial}(\boldsymbol{\alpha}_{i\cdot}),\\
     	\boldsymbol{h}_{i\cdot}|B,\Theta^{(1)},\Theta^{(2)},\lambda\sim \mathcal{N}\left(\left( \widetilde{\boldsymbol{t}}_{i\cdot}B\right) ^{T}, c_{i}\Sigma^{(2)}+(1-c_{i})\Sigma^{(1)} \right) ,\\
     	\theta_{jl}^{c_{i}}\sim\text{Laplace}(0,\lambda),\\
     	\theta_{jj}^{c_{i}}\sim\text{Exponential}\left(\frac{\lambda}{2}\right) ,\\
     	\lambda\sim\text{Exponential}\left(\hat{\lambda}_{init}\right),\\
     	B_{kj}\sim N(0,10000),
     	\end{gathered}
     \end{equation}
for $ i=1,\dots,N $, $ 1\leq l<j\leq p $, $ k=1,\dots,K+1 $, and $ c_{i}\in\{0,1\} $.  Here $ \widetilde{\boldsymbol{t}}_{i\cdot} $ is the $ i $-th row of the $ N\times(K+1) $ design matrix $ \widetilde{T} $, the first column of which is $ \boldsymbol{1}_{N} $. $ B_{(K+1)\times (p-1)} $ is the coefficient matrix showing how the covariates in $ \widetilde{T} $ relate to the alr transformed relative abundance.  Moreover,  $\hat{\lambda}_{init}=\mathop{\arg\max}_{\lambda>0}p\left( \lambda|\widehat{\Sigma}^{(1)},\widehat{\Sigma}^{(2)}\right)$ and non-informative normal priors are added on the elements of $ B $.
Then the joint posterior distribution $ p(H,B,\Theta^{(1)},\Theta^{(2)},\lambda|Y) $ can be solved, where the $ i $-th row of $ H $ is $ \boldsymbol{h}_{i\cdot} $. MDiNE applies Cholesky decompositions to precision matrices $ \Theta^{(k)},k=1,2 $ and uses Hamiltonian Monte Carlo (HMC) to estimate parameters by calculating the posterior mean.

MDiNE makes interval estimates for both parameters of the network and the difference between networks available. However, MDiNE has no zero-inflation parameter, so it may encounter problems in controlling data with excessive zeros. Besides, MCMC sampling may take too much time and increase computational burden. MDiNE uses a binary covaraite to define two networks. Thus, it cannot explore how networks change with a continuous covariate now, but this can be further studied.

\subsection{Differential Markov random field analysis}
Since previous graphical models cannot assign statistical significance or uncertainties for estimated associations, \citep{MarRanField2019} proposes a Markov random field model and a hypothesis testing framework to detect microbial network structure and differences between networks.

To conduct Markov network analysis, \citep{MarRanField2019} first obtains binary measurements $ \boldsymbol{\tilde{x}}_{i\cdot},i=1,\dots,n $ based on relative abundance. After that, the conditional dependencies among elements of $ \boldsymbol{\tilde{x}}_{i\cdot}$ can be characterized by a binary Markov random field. The joint distribution of each piece of data is $p\left(\boldsymbol{\tilde{x}}_{i\cdot}\right) \propto \exp\left(\sum_{(j,l)\in E}\theta_{jl}\tilde{x}_{ij}\tilde{x}_{il} \right)$ where $E$ is the edge set corresponding to conditional dependencies. Here, $ \theta_{jl} $ quantifies the conditional dependency of taxon $ j $ and $ l $, and a positive value indicates that two taxa co-occur, while a negative value means co-exclusiveness. 
    
Suppose that $ \widetilde{X}_{n_{k}\times p}^{(k)} $ is the binary data of the $ k $-th group and that $ \Theta^{(k)}=(\theta_{jl}^{(k)})_{p\times p} $ is the matrix able to capture the conditional dependency structure of the $ k $-th situation, $ k=1,2 $. \citep{MarRanField2019} then conducts a global test to identify whether $\Theta^{(1)}$ and $\Theta^{(2)}$ are the same. Since these two networks may only differ by a small number of links, the sparsity of the differential network can be leveraged, and the null hypothesis is equivalent to:
    \begin{equation}
    	H_{0}:\max_{1\leq j<l\leq p}\left|\theta_{jl}^{(1)}-\theta_{jl}^{(2)}\right|=0.
    	\label{eq:MarRanField_globalH0}
    \end{equation} 
    
If the null hypothesis in Equation~\ref{eq:MarRanField_globalH0} is rejected, then \citep{MarRanField2019} performs multiple testing to check the existence of entrywise changes:
    \begin{equation}
    	H_{0,jl}:\theta_{jl}^{(1)}=\theta_{jl}^{(2)} \quad \text{v.s.} \quad H_{1,jl}:\theta_{jl}^{(1)}\neq\theta_{jl}^{(2)},\  \forall  1\leq j<l\leq p.
    	\label{eq:MarRanField_entryTest}
    \end{equation}
Denote the standardized entrywise difference test statistic of Equation~\ref{eq:MarRanField_entryTest} as $ D_{jl} $. If $ |D_{jl}| $ is larger than the threshold $ \tau >0 $, then the null hypothesis in Equation~\ref{eq:MarRanField_entryTest} is rejected. 
        
Denote $ \boldsymbol{\widetilde{x}}_{i,-j}^{(k)}=(\widetilde{x}_{i1}^{(k)},\dots,\widetilde{x}_{i,j-1}^{(k)},\widetilde{x}_{i,j+1}^{(k)},\dots,\widetilde{x}_{ip}^{(k)}), 1\leq i\leq n_k $. Since we can obtain the conditional distribution of $ \widetilde{x}_{ij}^{(k)} $ given $ \boldsymbol{\widetilde{x}}_{i,-j}^{(k)} $, $ \theta_{jl}^{(k)}$ can be estimated \textit{via} a $ \ell_{1} $ penalized nodewise logistic regression according to \citep{Ravikumar2010}. However, estimators $ \hat{\theta}_{jl}^{(k)}$ obtained this way are biased, and \citep{MarRanField2019} uses a nonlinear regression to correct for the bias with residual error denoted as $\epsilon_{ij}^{(k)}$. \citep{MarRanField2019} follows \citep{Zhang2014} to project the residual onto the direction of a score vector $ \boldsymbol{v}_{jl}^{(k)}$ so as to obtain $\check{\theta}_{jl}^{(k)}$.
However, $ \check{\theta}_{j,l}^{(k)} $ may have different variance needing adjustment. Let $v_{i,j,l}^{(k)}$ be the calculated oracle score vector based on $\Theta^{k}$; then, as pointed out in \citep{MarRanField2019}, $ \check{\theta}_{j,l}^{(k)} $ is expected to be close to $ \tilde{\theta}_{j,l}^{(k)}=\theta_{j,l}^{(k)}+\frac{1}{n_{k}}\sum_{i=1}^{n_{k}}\frac{2v_{i,j,l}^{(k),o}\epsilon_{ij}^{(k)}}{\boldsymbol{q}_{jl}^{(k)}} $ with specified ${\boldsymbol{q}_{jl}^{(k)}}$ for the good initial estimator $ \hat{\theta}_{j,l}^{(k)} $. Since the sample version of variance $\check{\boldsymbol{s}}_{jl}^{(k)}$ that approximates $ \text{var}(\check{\theta}_{j,l}^{(k)}) $ can be defined, the standardized entrywise difference used for Equation~\ref{eq:MarRanField_entryTest} is $D_{jl}=\frac{\check{\theta}_{jl}^{(1)}-\check{\theta}_{jl}^{(2)}}{\sqrt{\check{\boldsymbol{s}}_{jl}^{(1)}/n_{1} +\check{\boldsymbol{s}}_{jl}^{(2)}/n_{2}}},\quad 1\leq j<l\leq p.$
Then, the statistic for testing the null hypothesis in Equation~\ref{eq:MarRanField_globalH0} is defined as $M_{n,p}=\max_{1\leq j<l\leq p}D_{jl}^{2}$.
The threshold $ \tau $ can then be chosen by the false discovery proportion.

The model in \citep{MarRanField2019} is free of distribution assumptions on the data and can be used to interpret the co-occurrence and co-exclusiveness of the microbial communities naturally. It can be extended to study the case in which each node can take more than two discrete values. In this model, taking different cut-off values to transform raw data into binary data may cause different results. Moreover, the consistency of results needs to be further examined, such as results when dichotomizing the data in different ways.

\section{Discussion}
\subsection{Ways to handle environmental confounding factors}
Just like mining biological patterns as prior information considered in MPlasso \citep{MPLasso2017}, environmental factors, as special prior information, also have an important impact on microbial network inference. Environmental factors, such as pH, humidity, temperature, and soil type, can strongly affect the composition of the microbial community.  Understanding interactions among microbes and between microbes and environmental factors is of great interest. However, whether the edge in the microbial network is caused by an environmental factor or a direct interaction between two taxa is hard to distinguish, and many detected associations between microbes may be spurious without considering environmental factors \citep{mLDM2017}.

Microbial network inference can handle environmental factors by using environmental factors as additional species and calculating their association with microbes, or treating them as additional variables in regression and inferring the association of remaining abundances unaffected by the environment \citep{faust2012microbial}, such as mLDM \citep{mLDM2017} and kLDM \citep{kLDM2020}. In addition, as mentioned in \citep{faust2021open}, we can cluster samples according to key environmental factors and construct networks for each sample group separately in order to decrease environmental variation within a group.

Multiple ways are available to predict associations between species and environmental factors, but which one to use relies on concrete goals during the research. For instance, treating environmental factors as additional nodes helps us realize how they influence microbial community composition, while stratifying samples or regressing out environmental confounders is beneficial for inferring microbial interactions. However, the problem lies in establishing a reliable evaluation system of various treatments in order to properly manage environmental factors \citep{faust2021open}. 

\subsection{Ways to detect complex relationships}
As we have mentioned above, methods that estimate interaction by means of Pearson correlations \citep{SparCC2012}, covariance matrix or precision matrix all assume that the microbial correlation relationship is pairwise linear, which is not always the case. Sometimes nonlinear complex interactions may occur in the microbial community. An interaction among some species could be influenced by an additional one \citep{faust2021open}, and one species could be affected by other species \citep{faust2012microbial}. These cases could not be handled correctly with direct pairwise interaction detection methods mentioned in this article. However, high order interactions were shown to affect the microbial community both in simulations \citep{bairey2016high} and in experiments \citep{gould2018microbiome}. Hence, it is necessary to find ways to manage complex relationships, which could provide evidence for non-random co-association patterns \citep{cardona2016network}.

We can detect complex relationships by only considering simple relationships first. After the model is constructed, the biases between model estimations and actual microbial community behaviors may indicate the presence of complex relationships \citep{faust2021open}. However, the accuracy of this method remains doubtful, because, for example, an improper variable or wrong parameter could lead to the skewing of model predictions \citep{billick1994higher}. 

We can also handle complex relationship inferences by association rule mining, which involves enumerating all logical rules in the presence-absence of data in order to find valuable rules \citep{faust2012microbial}. For instance, if species $A$ exists only in the presence of both species $B$ and $C$, then neither the interaction between $A$ and $B$ nor $A$ and $C$ exists unless a third species appears. This can be regarded as an extreme example of relationship correction \citep{faust2021open}. People have reported several association rules containing more than two taxa in the past. However, possible reasons behind relationships detected may include true complex relationships and confounding results such as assembly of pairwise interactions and overfitting \citep{faust2021open}. Thus, it's still a problem to uncover and propose concrete reasons that can lead to the detection of complex relationships, while still avoiding meaningless results.

Another approach worth trying is multiple regression, which predicts the abundance of a species through incorporating the abundance of other species \citep{faust2012microbial}. The principle of multiple regression is simple and clear and is, therefore, widely applied. However, this method is limited by the difficulty of discovering the implication of results when we sort factors through automatic feature selection methods like sparse restriction. In this way, although we may be able to provide a roughly accurate prediction of the abundance of the species mathematically \citep{faust2012microbial}, the existence of causality in biology is doubtful. 

In addition, people also attempt to capture complex forms of ecological interactions through the mutual information criterion which measures the additional information known about one variable when the other variable is given \citep{butte1999mutual}. Maximum Information Coefficient (MIC) is widely used to capture novel correlations on large datasets \citep{cardona2016network}, which puts mutual information into a continuous distribution and identifies diverse interactions between variables, both linear and non-linear \citep{reshef2011detecting}. However, sample-size must be large enough to calculate the value of MIC accurately, while this can increase the difficulty of implementation \citep{li2016predicting}. In addition, recent research has questioned the equitability and statistical performance of MIC \citep{gorfine2012comment}.

Finally, a microbial network involves complex relationships that likely consist of edges jointing more than two nodes. Consequently, it is not an easy task to visualize them \citep{faust2012microbial}. Hypergraphs should be taken into consideration, and how to investigate and explain them needs further exploration \citep{faust2021open}.

\subsection{Trade-off between model accuracy and computational complexity}
We realize the fact that simple models may fail to fit the microbial community well, while complex models may lead to computational burden. For instance, compared with Pearson correlation, SparCC is proven to be more suitable for avoiding false interactions, while inducing higher computational complexity \citep{matchado2021network}. What’s more, correlation-based methods usually cannot distinguish direct from indirect interactions. In order to solve this problem, a large number of methods based on conditional dependence have been developed. Compared with correlation-based methods, some conditional dependence methods have a sharp increase in computational complexity which results in unsustainable running times \citep{matchado2021network}, such as hierarchical Bayesian models like mLDM \citep{mLDM2017} and kLDM \citep{kLDM2020}. Therefore, some tradeoff between model accuracy and computational complexity is another point we should take into account.

To solve this dilemma, we should carefully choose the analytic method to use considering the computational burden brought by concrete datasets and available computation resources. We should also consider whether it is feasible for the microbiome to be analyzed on a higher taxonomic level.

Since different methods can detect different edges in the same data, \citep{Weiss2016} assumes that model accuracy can be improved by using a combination of various analytical tools. As a matter of fact, some studies have reported experiments on linearly modeled data with foregone engineered interactions \citep{Weiss2016}. Compared with most individual methods, it has been shown that the accuracy of the ensemble method is significantly improved at the cost of sensitivity. Hence, when the given research requires us to obtain results of high precision, for instance, utilizing hypotheses of microbial interactions in order to identify co-culturing, integrated methods should be applied.

Finally, while engaged in evolving computational methods of microbial communities, we should not neglect matching them with current experimental research. In this way, we can verify our discoveries and finally try to elucidate the complex mechanism of microbial interactions \citep{matchado2021network}. 

\subsection{Lack of universal standardized simulation and evaluation}
Simulated datasets are often needed to test the performance of methods. Various ways are currently available for simulation. One famous approach is generating data from distributions, such as the logistic normal distribution adopted by CCLasso \citep{CCLasso2015}, COAT \cite{COAT2019} and gCoda \citep{gCoda2017}. Another widely used way is adopting models, especially differential equations, such as the famous generalized Lotka–Volterra (GLV) model \citep{Buffie2015} \citep{Bucci2016} \citep{Venturelli2018}. However, we are still confounded by the lack of a universal standard for simulating microbial community with ground-truth correlations. This makes it difficult to evaluate microbial network inference methods in a unified way since the performance of the proposed methods is largely affected by the data generation approach. If the artificial data are generated in a way that approximates assumptions of the microbial correlation network inference model, then the model would gain a relatively good performance. That is, one way of simulating data would favor methods similar to its generation approach \citep{faust2021open}. More importantly, the true generation process of microbial data in reality is unknown, which may be more complex than current simulation approaches. 

To avoid the evaluation bias caused by a single simulation method and judge the performance of various methods more objectively, it's necessary to adopt different simulation methods when testing. It is also important to separate developing new microbiome correlation network inference methods from comparison studies to ensure neutrality \citep{faust2021open}. Weiss et al. \citep{Weiss2016} compare the performance of eight correlation techniques, as well as their ability of handling the issues in compositional data. Hirano and Takemoto \citep{HiranoTakemoto2019} include more recent methods, such as REBACCA \citep{REBACCA2015} and CCLasso \citep{CCLasso2015}. Comparative studies show that correlation estimation methods based on compositional data do not always outperform classical methods, such as Spearman correlation-based and maximal information coefficient-based methods. This may result from the fact that newly proposed methods are developed under certain conditions or for particular problems.

\subsection{Lack of benchmark microbial network dataset}
In addition to simulated datasets, real datasets with verified biological microbial associations are also vital to evaluate those interaction inference methods. Biological gold standard network datasets for interaction inference are considered those datasets that have available sequencing data and validated known interactions. Lima-Mendez et al. \citep{LimaMendez2015} built a literature-curated collection of 574 interactions in marine eukaryotic plankton, in which both mutualism and parasitism interactions are included. Besides, Durán et al. \citep{Duran2018} examined root-associated microbial communities and detected some microbial interactions which have been verified experimentally. However, ground-truth datasets for microbiome correlation network inference are still scarce.

Several obstacles interfere with building benchmark datasets from real data. First, some interactions that happen in the real world may not be observed at all. Thus, the verified interactions are inadequate and incomplete. Using these datasets as benchmarks to test microbiome correlation network inference methods would improve our evaluation of how these methods discover known associations. Even if a method could catch sight of underlying interactions between two taxa, such interactions may be viewed as false since they were not observed experimentally. 

Second, as previously pointed out, interactions between taxa can be direct or indirect. In reality, they are complex, and it's difficult for some methods to tell which kind of interactions are inferred. Especially, correlation network inference methods can obtain direct interactions or indirect ones, such as high-order interactions where associations among some species are altered by an additional species. The gap between inferred interactions and expected ones brings a challenge to the development of benchmark microbial datasets. 

\subsection{Challenges in network inference using multi-omics data}
Nowadays, high-throughput omics techniques have improved enough to promote microbiome studies in providing large-scale data of high quality \citep{DiBellaJM2013}. Microbiome multi-omics datasets can be built by collecting multiple types of biological data, such as metagenomics, metatranscriptomics, 16S, mass spectrometry-based metaproteomics and metabolomics \citep{Helbling2012}. For example, the Integrative Human Microbiome Project (iHMP), focusing on the interactions between microbiome and host, provides multi-omics data from human host and microbiome \citep{iHMP2014}\citep{iHMP2019}. Different types of biological data provide information from different views, so consolidating them together can bring a more comprehensive view. However, in reality, it's rare that all samples have all types of features measured, owing to the difficulty of gathering multi-omics data \citep{Jiang2019}. For example, longitudinal multi-omics studies are conducted by following 132 objects for one year \citep{LloydPrice2019}. Therefore, the number of samples which can be used by the multi-omics method may not be as large as would otherwise be desired. Integration methods should be able to manage data with missing features and not rely on paired multiomics data. 

Moreover, when compiling multiple types of microbiome data, their specific data features should be considered first. For instance, OTU data are compositional, high-dimensional, sparse and heterogeneous. This is an obstacle when mixing multiomics methods from other disciplines with microbiome multiomics data. When combining such methods, random effects from each type of data need to be removed, especially for regression-based methods. Even more complexity is added by the possible associations among different omics layers \citep{Liu2021}. First separating omics analysis and then integrating them would miss such associations.
 
Integration of multi-omics data can help in the discovery of correlations among microbes, as well as achieving a more comprehensive understanding of environmental or other factors. To promote further research on this problem, both experimental techniques and statistical methods, as well as deep learning methods, need to be improved in the future.

\subsection{Extension in capturing dynamic microbiome interactions} 
Although microbial communities can remain stable under some suitable conditions \citep{Faith2013}, the composition of microbiomes can vary as time goes on since environmental conditions or other factors affecting them change. Time-series analysis studying how microbiome interactions change is of significant biological importance, such as in identifying community behavior. Several network inference methods have focused on temporal changes in microbiomes. Local similarity analysis (LSA) \citep{LSA2006} is one of these methods which gains insight into dynamic changes by identifying associations among species and those between species and environmental factors. It adopts dynamic programming to capture such changes in time series and infers associations based on a similarity score. Eiler et al. \citep{Eiler2012} adopted LSA to reveal contemporaneous and time-lagged correlation patterns among populations in community members of bacterioplankton and associations with environmental factors. LSA was further extended to eLSA \citep{eLSA2011} by considering time-series data with replicates, and the theoretical approach to approximate the statistical significance of local similarity analysis is proposed in \citep{eLSA2013}. 

Bayesian network models are also approaches widely adopted to explore dynamic changes in microbiome data in addition to LSA. These models can be divided into two subtypes: Dynamic Bayesian Networks (DBNs) and Temporal Event Networks (TENs). A dynamic Bayesian network uses a snapshot to represent the state of all variables at some time point and is composed of such snapshots \citep{Layeghifard2017}. McGeachie et al. \citep{McGeachie2016} constructed a DBN model to capture specific relationships and changes in microbial compositions of the infant gut microbial ecosystem. In a temporal event network, each node represents the time at which an event happens or changes. TENs are suitable for problems only involving a few changes in a given time range owing to their simplicity \citep{Layeghifard2017}.

Since nodes in a microbial network usually refer to taxa at different taxonomic levels, edges representing associations among taxa in the network have often gained little attention \citep{DeSmetMarchal2010}. Thus essential changes in these microbial associations may be left out. Recently, a Correlation-Centric Network (CCN) \citep{CCN2020} was introduced to study changes of correlations among microbial network members. In contrast to traditional networks, each node in the CCN represents a species–species correlation, and each edge represents the species shared by two correlations. Compared to the classical network, such network structure can better capture dynamic changes of correlations between taxa. 

\section{Conclusion}
Microbial interaction estimation provides an insight into the complex interplay that exists in microbial communities and the connections between the microbiome and its living environment. Statistical computational methods can infer such association networks and guide further biological studies. Early research used traditional correlation statistics, such as Pearson correlation, to estimate correlations among taxa without distinguishing between direct and indirect interactions. Conditional correlation network inference was further developed to focus on direct associations, and many methods were proposed under the assumption of only one underlying biological network behind microbiome compositional data. However, deeper studies relax this assumption by using mixture models. Of course, microbial communities can change with time or environmental factors. To compare interactions under different conditions, differential network inference methods have arisen. In this review, we follow the development of microbiome network inference and analyze typical methods of correlation networks, conditional correlation networks, mixture networks and differential networks. These methods tackle inference problems from various aspects, and they are based on different assumptions, even while facing some common challenges coming from the compositional data. One feature of compositional data is high-dimensionality, in which the number of samples may be much smaller than their features. This can make the covariance matrix degenerate and increase the difficulty of optimization. Furthermore, since OTU data can only provide relative abundances, the increase in absolute abundance of a single taxon will make relative abundances of other taxa decrease. Traditional correlation statistics can give spurious results as a consequence of such compositional bias. Different transformations are adopted to solve this problem including clr and alr transformations. Still, microbiome data are sparse, and large amounts of zeros can make trouble for these transformations. Specifically, zeros in the count data can be the result of truly absent and inefficient capture. Such fact is considered by some models. In addition to these problems, heterogeneity existing in microbiome data also challenges modeling and computation, but can be managed, to a certain extent, with the use of complex models, such as mixture networks. Meanwhile, computation burden needs to be weighed against accuracy.

After estimating correlations among taxa, their biological interpretation is further required, which is not an easy task. The value and sign of interactions estimated can provide a general guide for interpretation. More specific explanations require the support of biological experiments and the literature. As mentioned before, in addition to conditional correlation networks, most correlation networks do not distinguish between direct and indirect interactions. Thus, whether estimated association is caused by a third object or certain environmental factors needs to be checked. Another challenge for interpretation is that associations can be time-lagged \citep{FaustRaes2012}. That is to say, it takes time for the influence of one taxon or environmental factors on another taxon to appear. This makes interpretation of relationships mined more problematic and requires subsequent experimental validation.

In addition to types of networks analyzed in the main paper, microbiome network inference methods can be improved by accounting for more factors affecting the living microbiome, estimating more complex relationships and considering time-variable networks. How to make methods more scalable with less computational complexity, while, at the same time, guaranteeing efficiency, is also a task for future studies. As discussed before, some challenges in benchmarking real microbiome datasets need resolution in order to simulate reliable data for evaluation and propose universal valuation standards. All these problems require improvements in biological techniques and statistical methods together.

\section*{Acknowledgements}
This work was supported by the National Key Research and Development Program of China (No.2016YFA0502303), the National Key Basic Research Project of China (No. 2015CB910303), and the National Natural Science Foundation of China (No.31871342).
\vspace*{-12pt}

\nocite*{}
\bibliographystyle{plain}

\begin{landscape}
	\centering 
	\setlength{\tabcolsep}{1mm}{
		\begin{longtabu}{p{19mm}p{34mm}p{19mm}p{17mm}cp{36mm}p{34mm}} 
			\caption{Summary of microbiome network inference methodologies}
			\label{table1:summary}\\
			\toprule 
			{\textbf{Methods}} &
			{\textbf{\tabincell{c}{Approaches\\/Models}}} &
			{\textbf{\tabincell{c}{Preproc\\-essing}}} &
			{\textbf{Sparsity}} &
			{\textbf{\tabincell{l}{Compositional \\Bias Rectified}}}&
			{\textbf{Advantages}}&
			{\textbf{Drawbacks}}\\
			\hline
			\endfirsthead
			\multicolumn{7}{l}{(Continued from previous page)}                                                           \\
			\toprule 
			{\textbf{Methods}} &
			{\textbf{\tabincell{c}{Approaches\\/Models}}} &
			{\textbf{\tabincell{c}{Preproc\\-essing}}} &
			{\textbf{Sparsity}} &
			{\textbf{\tabincell{l}{Compositional \\Bias Rectified}}}&
			{\textbf{Advantages}}&
			{\textbf{Drawbacks}}\\
			\hline
			\endhead 
			\midrule 
			\multicolumn{7}{r}{(Continued on next page)} \\ 
            \endfoot 
            \bottomrule
			\endlastfoot

			\multicolumn{7}{l}{\textbf{Correlation network}}\\			
			\tabincell{l}{SparCC\\ (2012)} & 
			{$\bullet$ Pearson correlations \textit{via} Bayesian framework} 
			&Log-ratios 
			&\tabincell{l}{Thresh\\-olding}
			&TRUE 
			&{$\bullet$ Free of compositional bias} 
			&{$\bullet$ High computational complexity}\\
			
			\tabincell{l}{CCLasso\\ \ \ (2015)}                                 
			& $\bullet$ Latent variable model                                                              & Log-ratios                                
			& $\ell_1$ penalty   
			& TRUE                           
			& {$\bullet$ Guaranteeing the correlation matrix positive}                               
			& {$\bullet$ Nonlinear interaction undetected}      \\
			
			\tabincell{l}{REBACCA\\ \quad(2015)} 
			& 
			{$\bullet$ Linear system equivalent to the log-ratio transformations}
			&Log-ratios                                
			&$\ell_1$ penalty   
			&TRUE                           
			&{$\bullet$ Fitting with large sample size}                                             
			&{$\bullet$ High computational requirement}               \\
			\tabincell{l}{COAT\\ (2019)}                                      
			&{$\bullet$ Thresholding estimator for compositional data}                                     
			&CLR transform                             
			&Thresholding       
			&TRUE                           
			&{$\bullet$ Theoretical guarantees provided}                                            
			&{$\bullet$ Nonlinear interaction undetected}                                                  \\ 
			\midrule 
			\multicolumn{7}{l}{\textbf{Conditional correlation network}}\\
			\tabincell{l}{SPIEC-EASI\\  \quad(2015)}                  
			&{$\bullet$ Graphical models using selecting neighborhoods or inverse covariance} 
			& CLR transform                             
			& $\ell_1$ penalty   
			& TRUE                           
			&{$\bullet$ Compositional bias tackled}                                                 
			&{$\bullet$ Hard to recover networks having large hubs}                                        \\ 
			
			\tabincell{l}{gCoda\\ (2017)}                                     
			& {$\bullet$ Logistic normal distribution}                                                       
			&{No special treatment}                      
			& $\ell_1$ penalty   
			& TRUE                           
			& {$\bullet$ More computional efficient and stable than SPIEC-EASI}         
			& {$\bullet$ Non-convex optimization  function}                                                  \\ 
			\tabincell{l}{MPLasso\\ \ \ (2017)}                                   
			& {$\bullet$ Multivariate normal distribution with prior information incorporated}              
			& CLR transform                             
			& $\ell_1$ penalty   
			& TRUE                           
			& {$\bullet$ Prior knowledge incorporated}                                               
			&{$\bullet$ Untested on a dynamic model}                                                       \\ 
			\tabincell{l}{CD-Trace\\ \ \ (2019)}                                  
			&{$\bullet$ A loss function for compositional data with D-trace loss}                           
			&
			{No special treatment} 
			& $\ell_1$ penalty   
			& TRUE                           
			&{$\bullet$ Convex loss making computation easier than gCoda}                           
			& {$\bullet$ Heavy computation complexity}                                                      \\ 
			\tabincell{l}{CDTr\\ (2019)}                                      
			&{$\bullet$ A loss function for compositional data with D-trace loss}                         
			& 
			{No special treatment} 
			& $\ell_1$ penalty   
			& TRUE                           
			&{$\bullet$ Simpler form of loss function than that of CD-trace}                        
			& {$\bullet$ No theoretical guarantee of consistency}                                           \\ 
			\tabincell{l}{MInt\\ (2016)}                                      
			&{$\bullet$ Poisson-multivariate normal hierarchical model}                                     
			& {No special treatment}                      
			& $\ell_1$ penalty   
			& FALSE                          
			&{$\bullet$ Flexible mean-variance relationship}                                        
			&{$\bullet$ Compositional bias unrectified}                                                    \\ 
			\tabincell{l}{mLDM\\  (2017)}                                     
			& {$\bullet$ Lognormal-Dirichlet-Multinomial hierarchical model}                                 
			& {No special treatment}                     
			& $\ell_1$ penalty   
			& TRUE                           
			& {$\bullet$ Environmental factors incorporated}                                         
			&{$\bullet$ Heavy computation burden}                                                          \\ 	\tabincell{l}{HARMONIES\\\quad \ (2020)}                                 
			& {$\bullet$ Zero-inflated negative binomial distribution with Dirichlet process prior}     
			& {Model-based normalization}                 
			& $\ell_1$ penalty   
			& TRUE                           
			& {$\bullet$ Able to capture zeros and over-dispersion}                                  
			& {$\bullet$ Performance affected by small sample size}                                       \\ 
			\tabincell{l}{BC-GLASSO\\ \quad \ (2020)}                                 
			& {$\bullet$ Logistic normal multinomial distribution}                                        
			& ALR transform                             
			& $\ell_1$ penalty   
			& TRUE                           
			& {$\bullet$ Bias-corrected estimator proposed}                            
			& {$\bullet$ Potential interactions between the reference taxon and all other taxa not modeled} \\ 
			\tabincell{l}{CompGlasso\\ \quad \ (2020)}                                
			& {$\bullet$ Logistic normal multinomial distribution}                                         
			& ALR transform                             
			& $\ell_1$ penalty   
			& TRUE                           
			& {$\bullet$ Reference-invariant property}                      
			& {$\bullet$ Probably depending on the algorithm that is used}                                  \\ 
			\midrule
			\multicolumn{7}{l}{\textbf{Mixture network}}\\
			\tabincell{l}{MixMPLN\\ \quad (2019)}                                   
			& {$\bullet$ Mixture of multivariate poisson lognormal distribution}
			& {No special treatment}                      
			& $\ell_1$ penalty   
			& FALSE                          
			& {$\bullet$ Different microbial clusters considered}                                    
			& {$\bullet$ Heavy computation complexity}                                                      
			\\
			\tabincell{l}{MixMCMC\\\quad (2019)}                                   
			& {$\bullet$ Mixture of multivariate poisson lognormal distribution with MCMC sampling}         
			& {No special treatment}                        
			& $\ell_1$ penalty   
			& FALSE                          
			& {$\bullet$ Able to recover the topologies for graphs more accurately}                  
			& {$\bullet$ Significantly slower than MixMPLN}                                                 \\
			\tabincell{l}{MixGGM\\ \ \ (2019)}                                    
			& {$\bullet$ Mixture of multivariate Gaussian distribution}                                      
			& CLR transform                             
			& $\ell_1$ penalty   
			& TRUE                           
			& {$\bullet$ compositionality considered}                                                
			& {$\bullet$ Depending strongly on the conditional number of the inverse covariance matrix}     \\ 
			\tabincell{l}{kLDM\\ (2020)}                                      
			& {$\bullet$ K-Lognormal-Dirichlet-Multinomial hierarchical model}                               
			& {No special treatment}                      
			& $\ell_1$ penalty   
			& TRUE                           
			& {$\bullet$ Compositional bias and large variance of count data considered}             
			& {$\bullet$ Heavy computation burden}                                                          \\ 
			\midrule
			\multicolumn{7}{l}{\textbf{Differential network}}\\			
			\tabincell{l}{DCDTr\\ (2017)}                                     
			& {$\bullet$ A loss function for differential network estimation based on D-trace loss}          
			& 
			{No special treatment} 
			& $\ell_1$ penalty   
			& TRUE                           
			& {$\bullet$ Estimating the precision matrix difference directly}                        
			& {$\bullet$ Reasonability of exchangable condition remaining doubtful}                          \\ 
			\tabincell{l}{MDiNE\\ (2020)}                                     
			& {$\bullet$ Hierarchical Bayesian model}                                                        
			& ALR transform                             
			& $\ell_1$ penalty   
			& TRUE                           
			& {$\bullet$ Estimation of the differential network using a binary covariate considered} 
			& $\bullet$ Long running time                                                             \\	
			\tabincell{l}{Differential Mar\\-kov random \\field analysis\\\qquad(2019)} 
			& {$\bullet$ Markov random field model}                                                         
			& {Binary measurements of compositional data}
			& {Hypothesis testing} 
			& TRUE                           
			& {$\bullet$ Free of distribution assumptions}                                           
			& {$\bullet$ Strongly affected by the chosen discretization threshold}                          \\
		\end{longtabu}}
\end{landscape}

\end{document}